\documentclass[10pt,journal,compsoc]{IEEEtran}
\ifCLASSOPTIONcompsoc
  \usepackage[nocompress]{cite}
\else
  \usepackage{cite}
\fi
\ifCLASSINFOpdf
\else
\fi

\usepackage[nocompress]{cite}
\usepackage{subfigure} %luo, for subfigure
\usepackage{multirow} %luo, for table
\usepackage{textcomp} %luo
\usepackage[T1]{fontenc} %luo, for underscore symbol
\usepackage{amssymb} %luo
\usepackage{amsmath,bm} %luo
\usepackage{graphicx} %luo
\usepackage{epsfig} %luo
\usepackage{multirow} %shaowei, for table
\usepackage{flushend} %luo
\usepackage{color}

\usepackage{amsfonts}
\usepackage{url}
\usepackage{algorithmic}
\usepackage{algorithm}
\newcommand{\x}{\mathbf{x}}

\newtheorem{definition}{\textbf{Definition}}

\begin{document}
%
% paper title
% Titles are generally capitalized except for words such as a, an, and, as,
% at, but, by, for, in, nor, of, on, or, the, to and up, which are usually
% not capitalized unless they are the first or last word of the title.
% Linebreaks \\ can be used within to get better formatting as desired.
% Do not put math or special symbols in the title.
\title{Improved Consistent Weighted Sampling Revisited}
%
%
% author names and IEEE memberships
% note positions of commas and nonbreaking spaces ( ~ ) LaTeX will not break
% a structure at a ~ so this keeps an author's name from being broken across
% two lines.
% use \thanks{} to gain access to the first footnote area
% a separate \thanks must be used for each paragraph as LaTeX2e's \thanks
% was not built to handle multiple paragraphs
%
%
%\IEEEcompsocitemizethanks is a special \thanks that produces the bulleted
% lists the Computer Society journals use for "first footnote" author
% affiliations. Use \IEEEcompsocthanksitem which works much like \item
% for each affiliation group. When not in compsoc mode,
% \IEEEcompsocitemizethanks becomes like \thanks and
% \IEEEcompsocthanksitem becomes a line break with idention. This
% facilitates dual compilation, although admittedly the differences in the
% desired content of \author between the different types of papers makes a
% one-size-fits-all approach a daunting prospect. For instance, compsoc
% journal papers have the author affiliations above the "Manuscript
% received ..."  text while in non-compsoc journals this is reversed. Sigh.

\author{Wei Wu,
        Bin Li,
        Ling Chen,
        Chengqi Zhang, \emph{Senior Member, IEEE,}\\
        and Philip S. Yu, \emph{Fellow, IEEE}% <-this % stops a space
\thanks{W. Wu, L. Chen and C. Zhang are with Center for Artificial Intelligence, FEIT, University of Technology Sydney, Ultimo, NSW 2007, Australia. E-mail: william.third.wu@gmail.com; \{ling.chen,chengqi.zhang\}@uts.edu.au.}% <-this % stops a space
\thanks{B. Li is with Data61, CSIRO, Eveleigh NSW 2015, Australia. E-mail: bin.li@data61.csiro.au.}% <-this % stops a space
\thanks{P. Yu is with the CS Dept., University of Illinois at Chicago, Chicago, IL 60607, USA. E-mail: psyu@uic.edu.}}

\IEEEtitleabstractindextext{%
\begin{abstract}
Min-Hash is a popular technique for efficiently estimating the Jaccard similarity of binary sets. Consistent Weighted Sampling (CWS) generalizes the Min-Hash scheme to sketch weighted sets and has drawn increasing interest from the community. Due to its constant-time complexity independent of the values of the weights, Improved CWS (ICWS) is considered as the state-of-the-art CWS algorithm. In this paper, we revisit ICWS and analyze its underlying mechanism to show that there actually exists dependence between the two components of the hash-code produced by ICWS, which violates the condition of independence. To remedy the problem, we propose an Improved ICWS (I$^2$CWS) algorithm which not only shares the same theoretical computational complexity as ICWS but also abides by the required conditions of the CWS scheme. The experimental results on a number of synthetic data sets and real-world text data sets demonstrate that our I$^2$CWS algorithm can estimate the Jaccard similarity more accurately, and also compete with or outperform the compared methods, including ICWS, in classification and top-$K$ retrieval, after relieving the underlying dependence.
\end{abstract}

% Note that keywords are not normally used for peerreview papers.
}

% make the title area
\maketitle

% To allow for easy dual compilation without having to reenter the
% abstract/keywords data, the \IEEEtitleabstractindextext text will
% not be used in maketitle, but will appear (i.e., to be "transported")
% here as \IEEEdisplaynontitleabstractindextext when the compsoc
% or transmag modes are not selected <OR> if conference mode is selected
% - because all conference papers position the abstract like regular
% papers do.
\IEEEdisplaynontitleabstractindextext
% \IEEEdisplaynontitleabstractindextext has no effect when using
% compsoc or transmag under a non-conference mode.

% For peer review papers, you can put extra information on the cover
% page as needed:
% \ifCLASSOPTIONpeerreview
% \begin{center} \bfseries EDICS Category: 3-BBND \end{center}
% \fi
%
% For peerreview papers, this IEEEtran command inserts a page break and
% creates the second title. It will be ignored for other modes.
\IEEEpeerreviewmaketitle

\IEEEraisesectionheading{\section{Introduction}\label{sec:intro}}
% Computer Society journal (but not conference!) papers do something unusual
% with the very first section heading (almost always called "Introduction").
% They place it ABOVE the main text! IEEEtran.cls does not automatically do
% this for you, but you can achieve this effect with the provided
% \IEEEraisesectionheading{} command. Note the need to keep any \label that
% is to refer to the section immediately after \section in the above as
% \IEEEraisesectionheading puts \section within a raised box.

% The very first letter is a 2 line initial drop letter followed
% by the rest of the first word in caps (small caps for compsoc).
%
% form to use if the first word consists of a single letter:
% \IEEEPARstart{A}{demo} file is ....
%
% form to use if you need the single drop letter followed by
% normal text (unknown if ever used by the IEEE):
% \IEEEPARstart{A}{}demo file is ....
%
% Some journals put the first two words in caps:
% \IEEEPARstart{T}{his demo} file is ....
%
% Here we have the typical use of a "T" for an initial drop letter
% and "HIS" in caps to complete the first word.
\IEEEPARstart{N}{owadays}, data are growing explosively on the Web. In 2016, Google handled at least 2 trillion searches~\cite{google}; Facebook Messenger and Whatsapp handled 60 billion messages a day~\cite{facebook} -- Big data have been driving machine learning and data mining research in both academia and industry~\cite{dumbill2013revolution,rajaraman2012mining}. No matter how data mining and machine learning develop, in most tasks such as classification, clustering and retrieval, computing the similarity of data is one of the most fundamental operations. However, exact similarity computation has become daunting for big data due to the ``3V'' nature (volume, velocity and variety). For example, in the scenario of text mining, it is intractable to enumerate the complete feature set (e.g., over $10^8$ elements in the case of 5-grams in the original data~\cite{rajaraman2012mining}). Therefore, it is urgent to develop efficient yet accurate similarity estimation techniques.

A powerful solution to address the above challenge is to exploit Locality Sensitive Hashing (LSH) techniques~\cite{indyk1998approximate, gionis1999similarity}, which are tactfully designed to approximate certain similarity (or distance) measures. LSH adopts a family of hash functions to map similar objects to the same hash code with higher probability than dissimilar ones. Many LSH methods have been successfully developed, e.g., Min-Hash for the Jaccard similarity~\cite{broder1998min}, Sim-Hash for the angle-based distance~\cite{charikar2002similarity,manku2007detecting}, and LSH with $p$-stable distribution for the $l_p$ distance~\cite{datar2004locality}, among which Min-Hash has been verified particularly effective in document analysis based on the bag-of-words representation~\cite{shrivastava2014defense}. Recently, some variations of Min-Hash have further improved its efficiency~\cite{li2010b,li2012one,mitzenmacher2014efficient,shrivastava2014densifying}.

Min-Hash and its variations treat all the elements equally and select one element uniformly from the set. However, in many cases, one wishes to select an element with a probability in proportion to its importance (or weight). A typical scenario is the \emph{tf-idf} used in text mining, where each term is assigned to a positive value to indicate its importance for discriminating the documents in the corpus. Min-Hash cannot handle such weights properly. To address this limitation, weighted Min-Hash algorithms have been explored to approximate the generalized Jaccard similarity~\cite{haveliwala2000scalable}, which is used to measure the similarity of weighted sets. Existing works of weighted Min-Hash can be roughly classified into \emph{quantization-based} and \emph{sampling-based} approaches.

Quantization-based methods explicitly quantize each weighted element into a number of distinct and equal-sized subelements, which are treated \emph{independently} in the augmented universal set. Then the standard Min-Hash scheme is directly applied to the collection of subelements. The remaining float part of each weighted element resulting from the quantization can be handled by either simply rounding off or preserving with probability~\cite{haeupler2014consistent}. Obviously, the computational complexity of the quantization-based methods is proportional to the number of subelements. Such a computational cost is still unaffordable if there are numerous subelements.

To avoid computing hash values for all subelements, researchers have resorted to sampling-based methods. The pioneering work~\cite{gollapudi2006exploiting} introduces the notion of ``active indices'', which are \emph{independently} sampled on a weighted element from bottom to top, as a sequence of subelements whose hash values are monotonically decreasing. Since many inactive subelements are skipped, the computational complexity is reduced to be proportional to the logarithm of the weight~\cite{gollapudi2006exploiting}. Recently, Consistent Weighted Sampling (CWS)~\cite{manasse2010consistent}, Improved CWS (ICWS)~\cite{ioffe2010improved} and Practical CWS (PCWS) \cite{wu2017consistent} further reduce the computational complexity to be constant for each weighted element by considering only two active indices: The largest active index smaller than the weight of the $k$-th weighted element, denoted by $y_k$, and the smallest active index greater than the weight, denoted by $z_k$. So far, ICWS~\cite{ioffe2010improved} is recognized as the state-of-the-art algorithm for approximating the generalized Jaccard similarity. ICWS produces the hash code in the form of $(k_*,y_{k_*})$, where $k_*$ denotes the element obtaining the minimum hash value while $y_{k_*}$ denotes the largest active index sampled on the $k_*$-th weighted element. In~\cite{li20150}, the component $y_{k_*}$ in the hash code $(k_*,y_{k_*})$ is simply discarded because it empirically demonstrates that almost the same performance can be obtained by merely using $k_*$. Instead of uniformly discretizing the logarithm of the weight to generate $y_k$~\cite{ioffe2010improved}, Canonical CWS~\cite{wu2016canonical} considers uniformly discretizing the original weight to avoid the risk of violating the uniformity property of the CWS scheme.
%\footnote{In this paper we don't consider the setting knowing the upper bound of each weighted element in the corpus as~\cite{shrivastava2016simple}.}

As the CWS scheme generalizes the weighted Min-Hash scheme, it should satisfy the independence condition of the two components, $k_*$ and $y_{k_*}$, of the hash code. Unfortunately, we find that this condition does not hold in ICWS~\cite{ioffe2010improved} and its theoretical analysis is also questionable -- \emph{Therefore, ICWS does not comply with the CWS scheme}.

In order to address the above problem, in this paper we propose an Improved ICWS (I$^2$CWS) algorithm, which not only shares the same theoretical computational complexity as ICWS~\cite{ioffe2010improved} and satisfies the required independence condition, but also complies with the uniformity and consistency properties of the CWS scheme. To this end, I$^2$CWS samples $y_k$ and $z_k$ separately without deriving $z_k$ from $y_k$ as~\cite{ioffe2010improved} does, such that $k_*$ is finally independent of $y_{k_*}$. To validate that the proposed I$^2$CWS algorithm is able to estimate the generalized Jaccard similarity better than ICWS, we conduct comparative study on a number of synthetic data sets with different distributions to demonstrate the merit of I$^2$CWS as a more accurate estimator. In addition, we also conduct extensive empirical tests on a number of real-world text data sets to compare the proposed I$^2$CWS algorithm with the state-of-the-arts in classification and top-$K$ retrieval. The experimental results demonstrate that I$^2$CWS can not only estimate the generalized Jaccard similarity more accurately than ICWS, but also compete with or outperform the compared methods, including ICWS, after relieving the underlying dependence. In summary, our contributions are four-fold:
\begin{enumerate}
\item We revisit ICWS~\cite{ioffe2010improved} and show that this state-of-the-art actually violates the independence condition of the two components, $k_*$ and $y_{k_*}$, of the hash code; so ICWS does not comply with the CWS scheme.
\item We propose the I$^2$CWS algorithm, which not only complies with the CWS scheme but also has the same computational complexity as ICWS.
\item We conduct comprehensive comparative study of the ability of I$^2$CWS and ICWS in estimating the generalized Jaccard similarity and find that I$^2$CWS acquires an estimator with smaller errors than ICWS.
\item We observe some interesting findings about the CWS algorithms from the empirical study, which may be helpful for choosing or designing CWS algorithms.
\end{enumerate}

The remainder of the paper is organized as follows: Section~\ref{sec:prelimaries} briefly introduces the Min-Hash and CWS schemes. We review the state-of-the-art algorithm, ICWS~\cite{ioffe2010improved}, and point out its problems in Section~\ref{sec:icws}. Then, we present our algorithm and its theoretical analysis in Section~\ref{sec:cws}. The experimental results are presented in Section~\ref{sec:exp} and the related work is discussed in Section \ref{sec:related}. Finally, we conclude the paper in Section~\ref{sec:con}.

\section{Preliminaries}
\label{sec:prelimaries}

In this section, we first give some notations which will be used throughout the paper. Then we will introduce the Min-Hash scheme and the CWS scheme.

Given a universal set $\mathcal{U}$ and its subset $\mathcal{S}\subseteq \mathcal{U}$, if for any element $k \in \mathcal{S}$, its weight $S_k=1$ or $S_k=0$, then we call $\mathcal{S}$ a binary set; if for any element $k \in \mathcal{S}$, $S_k\ge 0$, then we call $\mathcal{S}$ a weighted set. For hashing a binary set $\mathcal{S}$, a Min-Hash scheme assigns a hash value to each element $S_k$, $h: k \mapsto v_k$. By contrast, for hashing a weighted set, there is a different form of hash function $h: (k,y_k) \mapsto v_{k,y}$, where $y_k \in [0,S_k]$. A random permutation (or sampling) process returns the first (or uniformly selected) $k$ from a binary set (or $(k,y_k)$ from a weighted set). If the set is sampled $D$ times, we will obtain a fingerprint with $D$ hash values.

\subsection{Min-Hash}
\begin{definition}[Min-Hash~\cite{broder1998min}]
  Given a universal set $\mathcal{U}$ and a subset $\mathcal{S}\subseteq \mathcal{U}$, Min-Hash is generated as follows: Assuming a set of $D$ hash functions (or $D$ random permutations), $\{\pi_d\}_{d=1}^D$, are applied to $\mathcal{U}$, the elements in $\mathcal{S}$ which have the minimum hash value in each hash function (or which are placed in the first position of each permutation), $\{\min(\pi_d(\mathcal{S}))\}_{d=1}^D$, would be the Min-Hashes of $\mathcal{S}$.
\label{minwise hashing}
\end{definition}

Min-Hash~\cite{broder1998min} is an approximate algorithm for computing the Jaccard similarity of two sets. It is proved that the probability of two sets, $\mathcal{S}$ and $\mathcal{T}$, to generate the same Min-Hash value (hash collision) exactly equals the Jaccard similarity of the two sets $J(\mathcal{S},\mathcal{T})$:
$$\Pr[\min(\pi_d(\mathcal{S}))=\min(\pi_d(\mathcal{T}))] = J(\mathcal{S},\mathcal{T})=\dfrac{|\mathcal{S} \cap \mathcal{T}|}{|\mathcal{S} \cup \mathcal{T}|}.$$
The Jaccard similarity is simple and effective in many applications, especially for document analysis based on the bag-of-words representations~\cite{shrivastava2014defense}.

We can see from the above Min-Hash scheme that all elements in $\mathcal{U}$ are treated equally since all elements can be mapped to the minimum hash value with equal probability. If the standard Min-Hash scheme is applied to sampling a weighted set, the weight, which indicates different importance of each element, has to be simply replaced with 1 or 0 -- This treatment will result in serious information loss.

\subsection{Consistent Weighted Sampling}
In most real-world scenarios, weighted sets are more commonly seen than binary sets. For example, a document is commonly represented as a \emph{tf-idf} set. In order to reasonably compute the similarity of two weighted sets, the generalized Jaccard similarity was introduced in~\cite{haveliwala2000scalable}. Considering two weighted sets, $\mathcal{S}$ and $\mathcal{T}$, the generalized Jaccard similarity is defined as
\begin{equation}
generalized J(\mathcal{S},\mathcal{T})=\dfrac{\sum_{k}\min(S_k,T_k)}{\sum_{k}\max(S_k,T_k)}.
\label{eq:gjaccard}
\end{equation}

In order to efficiently compute the generalized Jaccard similarity, the Consistent Weighted Sampling (CWS) scheme has been proposed in~\cite{manasse2010consistent}.
\begin{definition}[Consistent Weighted Sampling~\cite{manasse2010consistent}] \label{def:cws}
  Given a weighted set $\mathcal{S} = \{S_1,\ldots,S_n\}$, where $S_k\ge 0$ for $k\in\{1,\ldots,n\}$, Consistent Weighted Sampling (CWS) generates a sample $(k, y_k): 0 \le y_k \le S_{k}$, which is uniform and consistent.
\end{definition}
\begin{itemize}
\item \textbf{Uniformity:} The subelement $(k, y_k)$ should be uniformly sampled from $\bigcup_{k}(\{k\} \times [0,S_{k}])$, i.e., the probability of selecting the $k$-th element is proportional to $S_{k}$, and $y_k$ is uniformly distributed in $[0, S_{k}]$.
\item \textbf{Consistency:} Given two non-empty weighted sets, $\mathcal{S}$ and $\mathcal{T}$, if $\forall k, T_k \le S_k$, a subelement $(k, y_k)$
is selected from $\mathcal{S}$ and satisfies $y_k \le T_k$, then $(k, y_k)$ will also be selected from $\mathcal{T}$.
\end{itemize}
  CWS has the following property $$\Pr[{\rm CWS}(\mathcal{S})={\rm CWS}(\mathcal{T})] = generalized J(\mathcal{S},\mathcal{T}).$$

\section{Review of ICWS}
\label{sec:icws}

Based on the generalized Jaccard similarity, some efficient CWS algorithms have been proposed~\cite{manasse2010consistent,ioffe2010improved}. To the best of our knowledge, Improved Consistent Weighted Sampling (ICWS)~\cite{ioffe2010improved} is remarkable in both theory and practice, and considered as the state-of-the-art method for weighted Min-Hash~\cite{li20150}. In this section, we will briefly review the method derived in~\cite{ioffe2010improved} and point out its potential problems.

\subsection{Derivation of ICWS}
\label{subsec:icws}

ICWS achieves a constant computational complexity independent of the weight, $S_k$, as it only introduces two special active indices,
\begin{equation}\label{eq:yus}
  y_k = u_{k1}S_k,
\end{equation}
which is uniformly sampled in $[0, S_{k}]$ as the largest active index less than $S_k$, and
\begin{equation}\label{eq:zus}
  z_k=\dfrac{S_k}{u_{k2}},
\end{equation}
which is sampled as the smallest active index greater than $S_k$, where $u_{k1},u_{k2}\sim{\rm Uniform}(0,1)$. Now one has $0 \le y_k \le S_k \le z_k < +\infty$.

In order to make $y_k$ uniformly distributed in $[0,S_k]$, ICWS employs the following equation
\begin{equation}
  \ln y_k=\ln S_k - r_k b_k,
\label{uniform2}
\end{equation}
where $b_k \sim {\rm Uniform}(0,1)$ and $r_k \sim {\rm Gamma}(2,1)$. Eq.~(\ref{uniform2}) can be rewritten as $y_k = S_k \exp(-r_k b_k)$, where $\exp(-r_k b_k) \sim{\rm Uniform}(0,1)$ can be proved. The proof for the uniformity property of ICWS~\cite{ioffe2010improved} is based on Eq.~(\ref{uniform2}). Thus, $y_k$ is uniformly distributed in $[0,S_k]$.

In the algorithmic implementation of ICWS, the above equation is replaced with the following equation for the sake of consistency\footnote{Due to the floor function in Eq.~(\ref{uniform1}), ICWS can produce the same $y_k$ if the values of $S_k$ in different weighted sets have only a small difference, which makes the collision of $(k,y_k)$ possible.}
\begin{equation}
  \ln y_k= r_k\left(\left\lfloor \dfrac{\ln S_{k}}{r_k}+ \beta_{k} \right\rfloor - \beta_{k}\right),
\label{uniform1}
\end{equation}
where $\beta_k \sim {\rm Uniform}(0,1)$. It can be proved that $\ln y_k$ in Eq.~(\ref{uniform2}) and $\ln y_k$ in Eq.~(\ref{uniform1}) have the same uniform distribution in $[\ln S_k - r_k, \ln S_k]$.
%in~\cite{ioffe2010improved}\footnote{In fact, the simplest programming implementation to generate a random variable $x$ from ${\rm Gamma}(2,1)$ is to first sample two uniform random variables, $u_{1},u_{2} \sim {\rm Uniform}(0,1)$ and then conduct a transformation as $x = -\ln (u_{1}u_{2})$.}.

In order to sample $k$ in proportion to $S_k$, ICWS implicitly makes use of a nice property of the exponential distribution: If each hash value $a_{k'}$ of the $k'$-th element is drawn from an exponential distribution parameterized with the corresponding weight, i.e., $a_{k'} \sim {{\rm Exp}(S_{k'})}$, the minimum hash value $a_k$ will be sampled in proportion to $S_k$,
\begin{equation}
  \Pr[a_k=\min\{a_1,\ldots,a_n\}]=\dfrac{S_k}{\sum_{k'} S_{k'}}.
\label{minexp}
\end{equation}
According to the CWS scheme~\cite{manasse2010consistent}, ICWS also requires that $a_k$ and $y_k$ are mutually \emph{independent} such that ICWS should satisfy
\begin{equation}
  {\rm pdf}(y_k, a_k) = {\rm pdf}(y_k){\rm pdf}(a_k) = \dfrac{1}{S_k} \cdot (S_k\mathrm{e}^{-S_ka_k}),
\label{exp}
\end{equation}
which indicates $y_k\sim {\rm Uniform}(0,S_k)$ and $a_k\sim {\rm Exp}(S_k)$.

The uniform distribution of $y_k$ has been satisfied in Eq.~(\ref{uniform1}). To construct an exponential distribution for $a_k$, ICWS adopts the following equation:
\begin{equation}
a_k = \dfrac{c_k}{z_k} = \dfrac{c_k}{y_k \exp(r_k)}, \label{eq:ayvu}
\end{equation}
where $c_k \sim {\rm Gamma}(2,1)$. In~\cite{ioffe2010improved}, $a_k$ has been proved to follow the exponential distribution ${\rm Exp}(S_k)$.

Based on Eq.~(\ref{eq:ayvu}), the selected $k_*$-th element is returned via $k_* = \arg\min_k a_k$. Finally, a hash code $(k_*,y_{k_*})$ is produced through ICWS. The ICWS algorithm introduced in~\cite{ioffe2010improved} is summarized in Algorithm~\ref{alg:icws}.

%So far the derivation seems correct via the theoretical analysis in \cite{ioffe2010improved}.
%It first samples $y_k$ using Eq.~(\ref{uniform1}) and $z_k = y_k(u_{k1}u_{k2})^{-1}$ which derives from Eq.~(\ref{eq:rzy}).

\begin{algorithm}[t]
\fontsize{8pt}{\baselineskip}\selectfont{\caption{The ICWS Algorithm~\cite{ioffe2010improved}}
%\fontsize{8pt}{0.85\baselineskip}\selectfont{\caption{Gumbel}
\label{alg:icws}
\begin{algorithmic}[1]
  \REQUIRE $\mathcal{S} = \{S_1,\cdots,S_n\}$
  \ENSURE $(k_*, y_{k_*})$
  \FOR {$k=1,\ldots,n$}
       \STATE $r_k \sim {\rm Gamma}(2,1)$
       \STATE $\beta_k \sim {\rm Uniform}(0,1)$
       \STATE $c_k \sim {\rm Gamma}(2,1)$
  \ENDFOR
  \FOR {all $k$ such that $S_k > 0$}
       \STATE $\ln y_k = r_k\left(\left\lfloor \dfrac{\ln S_{k}}{r_k}+ \beta_{k} \right\rfloor - \beta_{k}\right)$
       \STATE $z_k = y_k \exp(r_k)$
        \STATE $a_k = \dfrac{c_k}{z_k}$
  \ENDFOR
  \STATE $k_* = \arg\min_{k}a_k$
  \RETURN $(k_*, y_{k_*})$
\end{algorithmic}
}
\end{algorithm}

\subsection{Issue: Dependence between $y_k$ and $a_k$}
\label{subsec:boi}

The derivation of ICWS briefed above seems reasonable. Unfortunately, we would point out that ICWS actually violates the independence condition of $y_k$ and $a_k$, which suggests that ICWS does not comply with the CWS scheme~\cite{manasse2010consistent}. In the following, we will show where the dependence in~\cite{ioffe2010improved} is and how the dependence is introduced.

In order to guarantee the global consistency of the active indices, that is, the same $y_k$ produces the same $z_k$, ICWS builds the relationship between $y_k$ and $z_k$ using the following equation:
\begin{equation}
  \ln z_k = \ln y_k + r_k.
\label{eq:rzy}
\end{equation}
By combining Eq.~(\ref{uniform2}) and Eq.~(\ref{eq:rzy}), we have
%\begin{eqnarray}
\begin{alignat}{3}
  y_k & = S_{k}(\exp(-r_{k}))^{b_{k}} &=& S_k(x_{1}x_{2})^{b_k}, \label{eq:ysub}\\
  z_k & = \dfrac{S_k}{(\exp(-r_{k}))^{1-b_{k}}} & =& \dfrac{S_k}{(x_{1}x_{2})^{1-{b_k}}}, \label{eq:zsub}
\end{alignat}
%\end{eqnarray}
where $r_k = -\ln(x_1 x_2)$ and $x_1,x_2 \sim {\rm Uniform}(0,1)$.

Eq.~(\ref{eq:ysub}) and Eq.~(\ref{eq:zsub}) seem very different from Eq.~(\ref{eq:yus}) and Eq.~(\ref{eq:zus}) which lay the foundation for the derivation of ICWS. However, we can easily prove $(x_{1}x_{2})^{b_k}, (x_{1}x_{2})^{1-{b_k}} \sim {\rm Uniform}(0,1)$ as follows: Let $x = x_{1}x_{2}$, we have ${\rm pdf}(x)= -\ln x, 0<x \leq 1$. Then, we let $m_{k1} = x^{b_k}$. By independence, ${\rm pdf}(x,{b_k})={\rm pdf}(x){\rm pdf}({b_k})= -\ln x$. We consider the transformation from $(x,{b_k})$ to $(m_{k1},w)$, where $m_{k1} = x^{b_k}$ and $w=x$, $0<w\leq m_{k1}\leq1$. By the Jacobian transformation, we have ${\rm pdf}(m_{k1},w) = {\rm pdf}(x, {b_k}) \left|\begin{matrix} {\rm det}\frac{\partial(x,{b_k})}{\partial(m_{k1},w)}\end{matrix}\right|= \frac{1}{m_{k1}}.$
Marginalizing out $w$ gives ${\rm pdf}(m_{k1}) = \int_{0}^{m_{k1}}{{\rm pdf}(m_{k1},w)}dw = 1,$ which indicates that $m_{k1}=(x_{1}x_{2})^{b_k} \sim {\rm Uniform}(0,1)$. Similarly, $m_{k2}= (x_{1}x_{2})^{1-{b_k}}\sim {\rm Uniform}(0,1)$ because of $1-{b_k}\sim {\rm Uniform}(0,1)$.

Therefore, Eq.~(\ref{eq:ysub}) and Eq.~(\ref{eq:zsub}) can be expressed in the following forms:
\begin{eqnarray}
  y_k  &=& S_k m_{k1} \label{eq:ysx}, \\
  z_k & =& \dfrac{S_k}{m_{k2}},\label{eq:zsx}
\end{eqnarray}
where $m_{k1}=(x_{1}x_{2})^{b_k} \sim {\rm Uniform}(0,1)$, $m_{k2}=(x_{1}x_{2})^{1-{b_k}} \sim {\rm Uniform}(0,1)$.

Eq.~(\ref{eq:ysx}) and Eq.~(\ref{eq:zsx}) are the actual distributions of $y_k$ and $z_k$ used in the derivation as well as the algorithmic implementation of ICWS~\cite{ioffe2010improved}, instead of the intended distributions based on the two \emph{independent} uniform random variables $u_{k1}$ and $u_{k2}$ used in Eq.~(\ref{eq:yus}) and Eq.~(\ref{eq:zus}). Now the problem has become obvious: \emph{$m_{k1}$ and $m_{k2}$ are dependent uniform variables, which suggests that $y_k$ and $z_k$ are dependent active indices -- This violates the independence condition of the active indices in the CWS scheme}.
%\cite{ioffe2010improved} provides a theoretical analysis of ICWS. However, we would like to challenge Eqs.~(\ref{eq:rzy}-\ref{eq:zsub}), which in fact share three basic uniform random variables, $u_{k1}, u_{k2}$ and $b_k$, will violate independence of $y_k$ and $a_k$, Eq.~(\ref{exp}).

In the above, we have uncovered the underlying dependence between $y_k$ and $z_k$, which further leads to the dependence between $y_k$ and $a_k$ according to Eq.~(\ref{eq:ayvu}). Therefore, in~\cite{ioffe2010improved} ${\rm pdf}(y_k, a_k) \neq {\rm pdf}(y_k){\rm pdf}(a_k)$, which is essentially contradictory to Eq.~(\ref{exp}) -- the basic assumption taken by~\cite{ioffe2010improved} in its theoretical analysis.

The remaining question is how the dependence between $y_k$ and $z_k$ is introduced? The origin is Eq.~(\ref{eq:rzy}), which cancels out $S_k$ to directly establish the relationship between $y_k$ and $z_k$. It seems that $z_k$ can be generated more easily using Eq.~(\ref{eq:rzy}); however, $z_k$ is independent from $y_k$ only conditioned on $S_k$.

\section{Improved ICWS}
\label{sec:cws}

In this section we propose a new algorithm for consistent weighted sampling, which utterly avoids the dependence problem stemming from Eqs.~(\ref{eq:rzy}-\ref{eq:zsub}) in ICWS~\cite{ioffe2010improved}. We also demonstrate that the proposed algorithm complies with the uniformity and consistency properties of the CWS scheme~\cite{manasse2010consistent}.

\subsection{The I$^2$CWS Algorithm}
\label{subsec:algo}

\begin{algorithm}[t]
\fontsize{8pt}{\baselineskip}\selectfont{\caption{The I$^2$CWS Algorithm}
%\fontsize{8pt}{0.85\baselineskip}\selectfont{\caption{Gumbel}
\label{alg:ccws}
\begin{algorithmic}[1]
  \REQUIRE $\mathcal{S} = \{S_1,\cdots,S_n\}$
  \ENSURE $(k_*, y_{k_*})$
  \FOR {$k=1,\ldots,n$}
       \STATE $r_{k1}, r_{k2}\sim {\rm Gamma}(2,1)$
       \STATE $\beta_{k1}, \beta_{k2} \sim {\rm Uniform}(0,1)$
       \STATE $c_k\sim {\rm Gamma}(2,1)$
  \ENDFOR
  \FOR {all $k$ such that $S_k > 0$}
       \STATE $t_{k2} = \left\lfloor \dfrac{\ln S_{k}}{r_{k2}}+\beta_{k2} \right\rfloor$
       \STATE $z_k = \exp (r_{k2}(t_{k2}-\beta_{k2}+1))$
       \STATE $a_k = \dfrac{c_k}{z_k}$
  \ENDFOR
  \STATE $k_* = \arg\min_{k}a_k$
  \STATE $t_{k_*1} = \left\lfloor \dfrac{\ln S_{k_*}}{r_{k_*1}}+\beta_{k_*1} \right\rfloor$
  \STATE $y_{k_*} = \exp (r_{k_*1}(t_{k_*1}-\beta_{k_*1}))$
  \RETURN $(k_*, y_{k_*})$
\end{algorithmic}
}
\end{algorithm}

To relieve the dependence between $y_k$ and $z_k$ (thus $y_k$ and $a_k$) and preserve the properties of the CWS scheme as well, we need to construct a CWS algorithm satisfying the following conditions: 1) $y_k$ is uniformly sampled from $[0,S_k]$; 2) $a_k$ complies with an exponential distribution parameterized with $S_k$; 3) $y_k$ is independent of $a_k$. To this end, we can completely abandon the shared random variables in Eq.~(\ref{eq:ysub}) and Eq.~(\ref{eq:zsub}) and directly consider Eq.~(\ref{eq:yus}) and Eq.~(\ref{eq:zus}):
\begin{enumerate}
  \item $z_k$ is independently sampled through $z_k=\frac{S_k}{u_{k2}}=\frac{S_k}{(\exp({-r_{k2}}))^{1-b_{k2}}}$, where $r_{k2} \sim {\rm Gamma}(2,1)$, $b_{k2} \sim {\rm Uniform}(0,1)$;
  \item The hash function is seeded with $z_k$ and outputs the hash value $a_k$ conforming to the exponential distribution with the parameter being $S_k$, that is, $a_k\sim {\rm Exp}(S_k)$ and obtain $k_* = \arg\min_k a_k$;
  \item $y_{k_*}$ is independently sampled through $y_{k_*}=S_{k_*}u_{k_*1}=S_{k_*}(\exp({-r_{k_* 1}}))^{b_{{k_*}1}}$, where $r_{k_* 1} \sim {\rm Gamma}(2,1)$, $b_{k_* 1} \sim {\rm Uniform}(0,1)$.
\end{enumerate}
Obviously, the above procedure not only preserves the uniformity of $(k,y_k)$ but also guarantees the independence between $y_k$ and $z_k$ (thus $y_k$ and $a_k$) because all the random variables for generating $y_k$ and $z_k$ are mutually independent.

On the other hand, in order to enable consistency, in the algorithmic implementation we follow ICWS~\cite{ioffe2010improved} to replace
\begin{eqnarray}
y_{k_*} & =& S_{k_*}(\exp({-r_{k_* 1}}))^{b_{{k_*}1}}, \label{eq:i2cwsy0} \\
z_k &=& \frac{S_k}{(\exp({-r_{k2}}))^{1-b_{k2}}} \label{eq:i2cwsz0}
\end{eqnarray}
with
\begin{eqnarray}
y_{k_*} &=& \exp \left(r_{k_*1}\left(\left\lfloor \frac{\ln S_{k_*}}{r_{k_*1}}+\beta_{k_*1} \right\rfloor-\beta_{k_*1}\right)\right), \label{eq:i2cwsy} \\
z_k &=& \exp \left(r_{k2}\left(\left\lfloor \frac{\ln S_{k}}{r_{k2}}+\beta_{k2} \right\rfloor-\beta_{k2}+1\right)\right), \label{eq:i2cwsz}
\end{eqnarray}
respectively. The two sampling equations for $y_{k_*}$, Eqs.~(\ref{eq:i2cwsy0}) and (\ref{eq:i2cwsy}),  share the same distribution: $\ln y_{k_*} \sim {\rm Uniform}(\ln S_{k_*} -r_{k_*1}, \ln S_{k_*}$); The two sampling equations for $z_k$, Eqs.~(\ref{eq:i2cwsz0}) and (\ref{eq:i2cwsz}), share the same distribution: $\ln z_k \sim {\rm Uniform}(\ln S_k, \ln S_k+r_{k2})$. In this way, we are able to independently acquire the same $y_k$ (and $z_k$) even if $S_k$ changes slightly. Our algorithm, which is named I$^2$CWS, is summarized in Algorithm~\ref{alg:ccws}.

\textbf{Computational Complexity}: It is worth noting that $y_{k_*}$ is only computed once after obtaining the index of the minimum hash value, $k_*$ (Lines 11--13 in Algorithm~\ref{alg:ccws}); while the for-loops (Lines 6--10) in both Algorithm~\ref{alg:icws} and Algorithm~\ref{alg:ccws} have the same complexity. All the random variables can be sampled off-line. Therefore, the I$^2$CWS algorithm shares the same computational complexity as ICWS~\cite{ioffe2010improved}.

\subsection{Analysis}

In this subsection, we will demonstrate that the proposed I$^2$CWS algorithm generates a sample $(k,y_k)$ satisfying the uniformity and consistency properties of the CWS scheme (see Definition~\ref{def:cws})~\cite{manasse2010consistent}.

\subsubsection{Uniformity}

In the following we drop the element index $k$ for conciseness. In the proof of uniformity, we adopt $y=S(\exp(-r_{1}))^{b_1}$ and $z=\frac{S}{(\exp(-r_{2}))^{1-b_{2}}}$, where $r_1, r_2\sim {\rm Gamma}(2,1)$, $b_1, b_2\sim {\rm Uniform}(0,1)$ due to the same distribution mentioned in Section \ref{subsec:algo}.

In Section~\ref{subsec:boi} we have shown that, let $u = (\exp(-r))^{b}$, $r\sim {\rm Gamma}(2,1)$, $b\sim {\rm Uniform}(0,1)$, then $u \sim {\rm Uniform}(0,1)$. Thus we have $y=S(\exp(-r_{1}))^{b_1} = Su_{1} \sim {\rm Uniform}(0,S)$ and $z=\frac{S}{(\exp(-r_{2}))^{1-b_{2}}} = \frac{S}{u_{2}}$, where $u_{1},u_{2} \sim {\rm Uniform}(0,1)$.

Next we will show that $a \sim {\rm Exp}(S)$ is also true. Combining $a = \frac{c}{z}$, where $c\sim {\rm Gamma}(2,1)$ (Line 9 in Algorithm~\ref{alg:ccws}) and $z = \frac{S}{u_{2}}$ (see above), we obtain $a = \frac{cu_2}{S}$. Let $l = cu_2$, we have ${\rm pdf}_{L}(l) = \int_{0^+}^1 \frac{1}{u_2}{\rm pdf}_{U_2}(u_2){\rm pdf}_{C}\left(\frac{l}{u_2}\right)du_2= \int_{0^+}^1 \frac{1}{u_2} \cdot 1 \cdot \frac{l}{u_2} \exp(-\frac{l}{u_2}) du_2  = \exp(-l)$, which indicates $l\sim {\rm Exp}(1)$. For $a=\frac{l}{S}$, through the Jocobian transformation, we have ${\rm pdf}_{A}(a)={\rm pdf}_{L}(l)|\frac{dl}{da}|=S\mathrm{e}^{-Sa}$, which further indicates $a\sim {\rm Exp}(S)$.
%Therefore, $y$ is uniformly distributed in $[0,S]$ and $a$ conforms to an exponential distribution parameterized with $S$.

For all the weights $\{S_1,\ldots,S_n\}$ in weighted set $\mathcal{S}$, there exist a set of exponential distributions parameterized with the corresponding weights. According to Eq.~(\ref{minexp}), $a_{k_*}$ is the minimum hash value with a probability in proportion to $S_{k_*}$, $\Pr(a_{k_*}=\min_{k} a_{k})=\frac{S_{k_*}}{\sum_{k} S_{k}}$. Note that $a$ is a function of $z$ which is sampled independently of $y$, so $a$ is independent of $y$. Consequently, we have ${\rm pdf}(y, a)= {\rm pdf}(y){\rm pdf}(a)$. Therefore, $(k_*, y_{k_*})$ is uniformly sampled from $\bigcup_{k}(\{k\} \times [0, S_{k}])$.

\subsubsection{Consistency}

In the following we will demonstrate that, for two non-empty weighted sets $\mathcal{S}$ and $\mathcal{T}$, if $\forall k, T_{k} \le S_{k}$, a subelement $(k_*, y_{k_*})$ is sampled from $\mathcal{S}$ and satisfies $y_{k_*} \le T_{k_*}$, then $(k_*, y_{k_*})$ will also be sampled from $\mathcal{T}$.

Considering the $k_*$-th element, we have $t_{k_*1}^{S} = \lfloor \frac{\ln S_{k_*}}{r_{k_*1}}+\beta_{k_*1} \rfloor$, thus $\frac{\ln S_{k_*}}{r_{k_*1}}+\beta_{k_*1}-1 < t_{k_*1}^{S} \le \frac{\ln S_{k_*}}{r_{k_*1}}+\beta_{k_*1}$. By hypothesis, $y_{k_*}^{S} = y_{k_*} \le T_{k_*}\le S_{k_*}$, then
$\frac{\ln T_{k_*}}{r_{k_*1}}+\beta_{k_*1}-1 < t_{k_*1}^{S} = \frac{\ln y_{k_*}}{r_{k_*1}}+ \beta_{k_*1}
\le \frac{\ln T_{k_*}}{r_{k_*1}}+ \beta_{k_*1}$. Obviously,
$$t_{k_*1}^{S} = \left\lfloor \frac{\ln T_{k_*}}{r_{k_*1}}+\beta_{k_*1} \right\rfloor = t_{k_*1}^{T},$$
which indicates $y_{k_*}^{S}=y_{k_*}=y_{k_*}^{T}$. Thus $y_{k_*}^{S}$ and $y_{k_*}^{T}$ will be sampled from the $k_*$-th elements of $\mathcal{S}$ and $\mathcal{T}$, respectively. Similarly, we can also show that any sample satisfying $y_{k_*}\le T_{k_*}\le S_{k_*}$ can be selected to obtain $t_{k_*2}^{S} = t_{k_*2}^{T}$, which indicates $z_{k_*}^{S}=z_{k_*}=z_{k_*}^{T}$.

On the other hand, we notice that, for any $k$, $a_k$ is essentially a monotonically non-increasing function of $S_k$:
\begin{eqnarray}
% \nonumber % Remove numbering (before each equation)
  a_k &=& \dfrac{c_k}{z_k} \nonumber\\
      &=& \dfrac{c_k}{\exp \left(r_{k2}\left(\left\lfloor \dfrac{\ln S_{k}}{r_{k2}}+\beta_{k2} \right\rfloor-\beta_{k2}+1\right)\right)}.\nonumber
\end{eqnarray}
Therefore, $\forall k, a_{k}^{T}\ge a_{k}^{S}$ due to $T_{k} \le S_{k}$, while $a_{k_*}^{T}=a_{k_*}^{S}=\min_{k} a_{k}^{S}$ because of $z_{k_*}^{S}=z_{k_*}^{T}$. As a result, $a_{k_*}^{T} = a_{k_*}^{S} \le a_{k}^{S} \le a_{k}^{T}$ and in turn $\arg\min_{k} a_{k}^{T}=\arg\min_{k} a_{k}^{S} = k_*$, which demonstrates that $(k_*,y_{k_*})$ is sampled from $\mathcal{S}$ and $\mathcal{T}$ simultaneously. Thus consistency holds.

In summary, I$^2$CWS strictly abides by the independence condition between $y_k$ and $a_k$, and also satisfies the uniformity and consistency properties of the CWS scheme.

\section{Experimental Results}
\label{sec:exp}

In the following, we first conduct comparative study on a number of synthetic data sets with different distributions to demonstrate that the proposed I$^2$CWS algorithm is able to estimate the generalized Jaccard similarity better than ICWS in Section \ref{subsec:error}. Then, we report the experimental results of the proposed I$^2$CWS algorithm and a number of state-of-the-art weighted Min-Hash and CWS algorithms on four real-world text data sets. We investigate the effectiveness and efficiency of the compared methods for classification in Section~\ref{subsec:classification} and for information retrieval in Section~\ref{subsec:retrieval}.

\subsection{Experimental Preliminaries}

\begin{table}[t]
\normalsize
\begin{center}
\fontsize{8pt}{1.2\baselineskip}\selectfont
\caption{\normalsize{Summary of the Data Sets}}
\label{tbl:datasets}
\begin{tabular}{|c|r|r|r|r|} \hline
\multirow{2}*{Data Set} &\# of  &\# of  & Average & Average Std \\
& Docs & Features & Density  & of Weights\\\hline
Syn0L10U & 1,000 & 100,000 & $0.005$ & 2.6362\\ \hline
Syn0L1U & 1,000 & 100,000 & $0.005$ & 0.2637\\ \hline
Syn0L0.1U & 1,000 & 100,000 & $0.005$ & 0.0264\\ \hline
Syn0L0.01U & 1,000 & 100,000 & $0.005$ & 0.0026\\ \hline
Syn0L0.001U & 1,000 & 100,000 & $0.005$ & 0.00026\\ \hline
Syn0L0.0001U & 1,000 & 100,000 & $0.005$ & 0.000026\\ \hline\hline
Syn2E1S & 1,000 & 100,000 & $0.005$ & 1.1962\\ \hline
Syn2E2S & 1,000 & 100,000 & $0.005$ & 2.3985\\ \hline
Syn2.5E1S & 1,000 & 100,000 & $0.005$ & 0.7233\\ \hline
Syn2.5E2S & 1,000 & 100,000 & $0.005$ & 1.4463\\ \hline
Syn3E1S & 1,000 & 100,000 & $0.005$ & 0.5189\\ \hline
Syn3E2S & 1,000 & 100,000 & $0.005$ & 1.0353\\ \hline
Syn3.5E1S & 1,000 & 100,000 & $0.005$ & 0.3997\\ \hline
Syn3.5E2S & 1,000 & 100,000 & $0.005$ & 0.8025\\ \hline
Syn4E1S & 1,000 & 100,000 & $0.005$ & 0.3223\\ \hline
Syn4E2S & 1,000 & 100,000 & $0.005$ & 0.6515\\ \hline\hline
Rcv1 & 20,000 & 47,236 & $0.0016$ & 0.0059\\\hline
Kdd & 20,000 & 20,216,830 & $0.0004$ & 0.0030\\\hline
Webspam & 350,000 & 16,609,143 & $0.0055$ & 0.0032\\\hline
Url & 2,396,130 & 3,231,961 & $0.000036$ & 0.0000027\\
%DBLP & 14,000 & 23,199 & $0.0032$ & 0.0027\\
\hline
\end{tabular}
\end{center}
\begin{flushleft}
\footnotesize{``\# of Docs'': size of the data set. ``\# of Features'': size of the dictionary (universal set) of the data set. ``Average Density'': ratio of the elements with positive weights to all the elements in the universal set (a small value indicates a sparse data set). ``Average Std of Weights'': standard deviation of the weights of the documents for each element (a large value indicates that the documents have very different weights for the corresponding element).}
\end{flushleft}
\end{table}

Seven state-of-the-art compared methods are used in our experiments:
\begin{enumerate}
\item \textbf{Min-Hash}: The standard Min-Hash scheme is applied by simply treating weighted sets as binary sets;
\item \textbf{WMH}: It applies Min-Hash to the collection of subelements which are generated by explicitly quantizing weighted sets and rounding the remaining float part of the subelements;
\item \textbf{[Haeupler et.~al.,~2014]~\cite{haeupler2014consistent}}: Compared to WMH, it preserves the remaining float part of the subelements with probability;
\item \textbf{[Gollapudi et.~al.,~2006]~\cite{gollapudi2006exploiting}}: It transforms weighted sets into binary sets by thresholding real-value weights with random samples and then applies the standard Min-Hash scheme (another algorithm is introduced in the same paper which improves WMH but extremely inefficient for real-value weights and thus not reported);
\item \textbf{ICWS~\cite{ioffe2010improved}}: It is introduced in Section~\ref{sec:icws}, which is currently the state-of-the-art for weighted Min-Hash in terms of both effectiveness and efficiency;
\item \textbf{[Li,~2015]~\cite{li20150}}: It approximates ICWS by simply discarding one of the two components, $y_k$, in the hash code $(k, y_k)$ of ICWS;
\item \textbf{CCWS~\cite{wu2016canonical}}: Instead of uniformly discretizing the logarithm of the weight as ICWS, it directly uniformly discretizes the original weight.
\end{enumerate}

\begin{figure}[t]
\centering
\includegraphics[width=\linewidth]{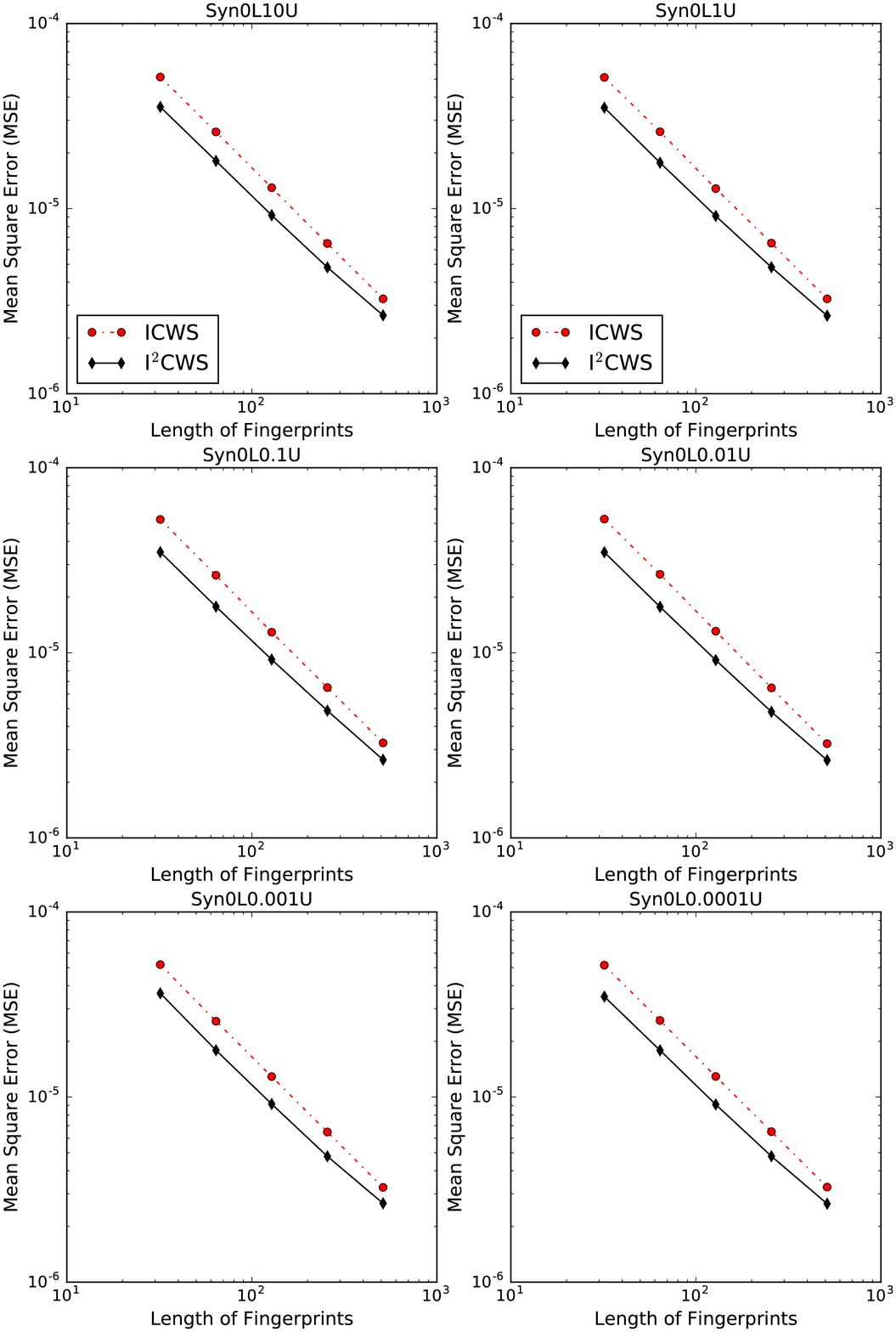}
\caption{Mean square errors (MSEs) on the synthetic weighted sets with uniform distributions. The $x$-axis denotes the length of fingerprints, $D$.}
\label{fig:uniform}
%\vspace{-10pt}
\end{figure}

All the compared algorithms are implemented in Matlab. We first apply all the algorithms to generate the fingerprints of the data. For WMH and [Haeupler et.~al.,~2014], each weight is scaled up by a factor of 10 for quantization of the subelements. Suppose that each algorithm generates $\x_S$ and $\x_T$, which are the fingerprints with the length of $D$ for the two real-value weighted sets, $\mathcal{S}$ and $\mathcal{T}$, respectively. The similarity between $\mathcal{S}$ and $\mathcal{T}$ is $$\mathbf{Sim}_{\mathcal{S},\mathcal{T}} = \sum_{d=1}^{D} \dfrac{\mathbf{1}(x_{S,d} = x_{T,d})}{D},$$ where $\mathbf{1}(state) = 1$ if $state$ is true, and $\mathbf{1}(state) = 0$ otherwise. The above equation calculates the ratio of the same Min-Hash values (i.e., collision) between $\x_S$ and $\x_T$, which is used to approximate the probability that $\mathcal{S}$ and $\mathcal{T}$ generate the same Min-Hash value, and to estimate the generalized Jaccard similarity. We set $D$, the parameter of the number of hash functions (or random samples), such that $D \in \{32,64,128,256,512\}$. All the random variables are globally generated at random in one sampling process. That is, the same elements in different weighted sets share the same set of random variables. All the experiments are conducted on a node of a Linux Cluster with $8 \times 3.1$ GHz Intel Xeon CPU (64 bit) and 1TB RAM.

\subsection{Results on Quality of Estimators}
\label{subsec:error}

\begin{figure}[t]
\centering
\includegraphics[width=0.945\linewidth]{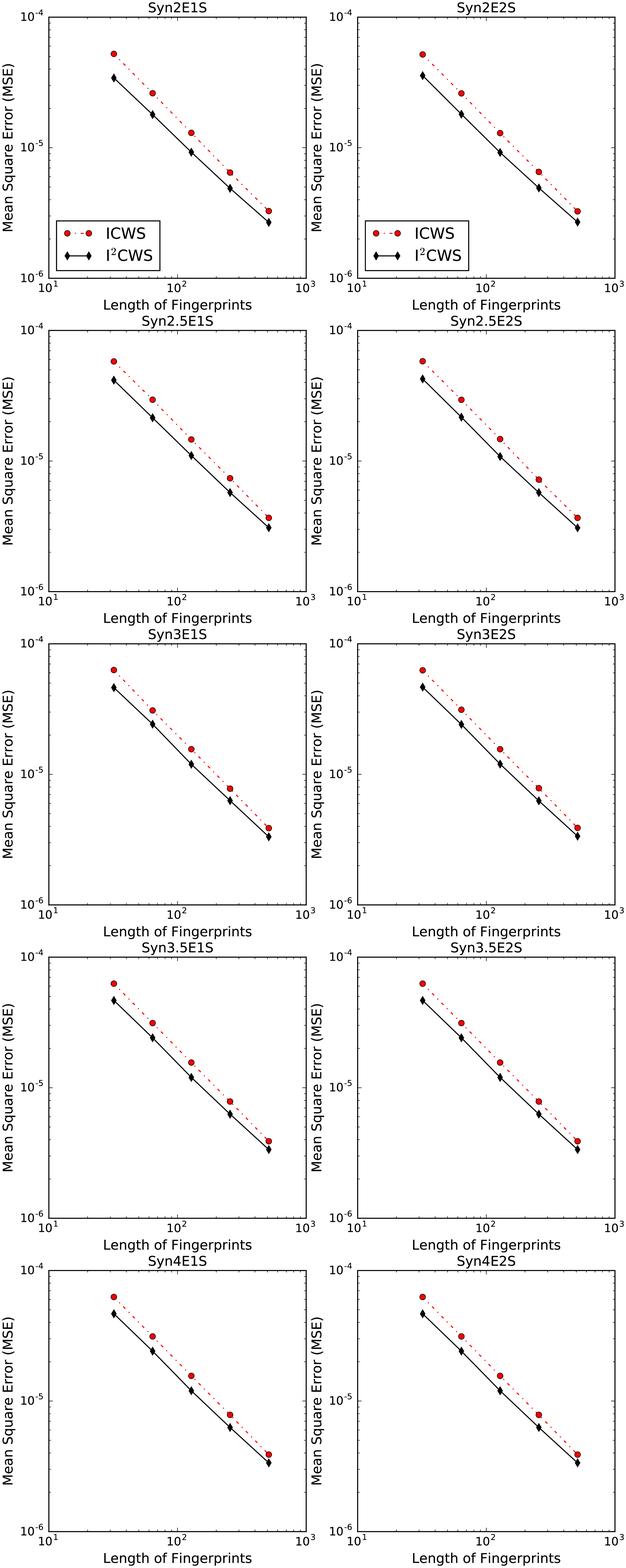}
\caption{Mean square errors (MSEs) on the synthetic weighted sets with powerlaw distributions. The $x$-axis denotes the length of fingerprints, $D$.}
\label{fig:powerlaw}
%\vspace{-10pt}
\end{figure}

In order to validate that the proposed I$^2$CWS algorithm is able to estimate the generalized Jaccard similarity better than the state-of-the-art algorithm, ICWS, we conduct the comparative study on a number of synthetic weighted sets with two different distributions of weights (see Table~\ref{tbl:datasets} for details) and present the empirical mean square errors (MSEs) of the estimators for the generalized Jaccard similarity by comparing the estimation result and the real generalized Jaccard similarity calculated using Eq.~(\ref{eq:gjaccard}).

\begin{enumerate}
\item \textbf{Syn$x$L$y$U}: Each data set of this group is a synthetic vector data set with 1,000 samples and 100,000 features. To generate 1,000 samples, we repeat the following procedure 1,000 times: First, we uniformly generate the dimensions where the values (i.e., weights) are non-zero. Second, the weights in the above dimensions are uniformly drawn from $[x,y]$. Finally we obtain a synthetic data set named Syn$x$L$y$U, where ``$x$L" indicates that the lower bound of the uniform distribution is $x$ and ``$y$U" indicates that the upper bound of the uniform distribution is $y$.
\item \textbf{Syn$x$E$y$S}: In order to simulate ``bag-of-words" in real text data following power-law distributions, we generate a second group of synthetic data sets, each of which contain 1,000 samples and 100,000 features as well. Similarly, we uniformly produce the dimensions, but the nonzero weights in each vector sample conform to a power-law distribution with the exponent parameter being $x$ and the scale parameter being $y$. After repeating 1,000 times, we obtain a data set dubbed Syn$x$E$y$S, where ``$x$E" indicates that the exponent parameter of the power-law distribution is $x$ and ``$y$S" indicates that the scale parameter of the power-law distribution is $y$.
\end{enumerate}

\subsubsection{Overall Results}

Figure \ref{fig:uniform} and Figure \ref{fig:powerlaw} show the comparison results in MSE for the estimation accuracy of generalized Jaccard similarity on two groups of synthetic data sets, respectively. Remarkably, the proposed I$^2$CWS algorithm clearly outperforms the state-of-the-art ICWS with smaller MSE in all cases. It is worth noting that the performance gain of I$^2$CWS is more obvious when the length of fingerprints is small, which implies that I$^2$CWS is more powerful in the scenarios where the computational and spatial budget is limited.

\subsubsection{Discussion on the Results}

The experimental results validate that our I$^2$CWS algorithm is able to approximate the generalized Jaccard similarity more accurately than ICWS. This phenomenon fundamentally justifies our preceding theoretical analysis in Section \ref{sec:icws} that ICWS essentially breaks the independence condition of $y_k$ and $a_k$ of the CWS scheme, while our I$^2$CWS algorithm utterly solves the problem. Consequently, our I$^2$CWS algorithm not only theoretically complies with the CWS scheme, but also empirically acquires the accurate estimator.

\subsection{Results on Classification}
\label{subsec:classification}

\begin{figure*}[ht]
\centering
\includegraphics[width=\linewidth]{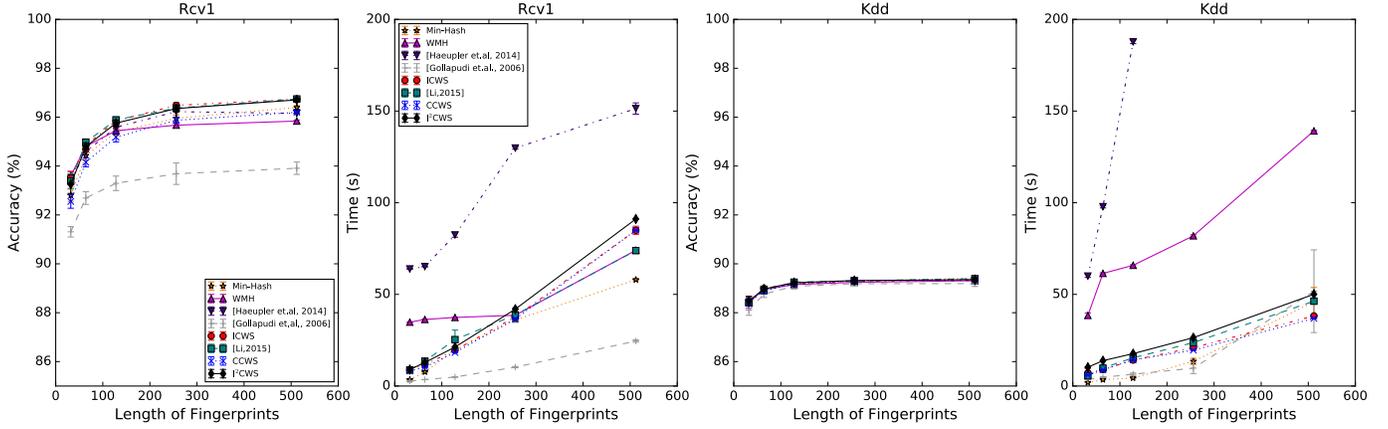}
\caption{Classification results in accuracy (odd columns) and runtime (even columns) of the compared methods on Rcv1 and Kdd. The $x$-axis denotes the length of fingerprints, $D$.}
\label{fig:classification}
%\vspace{-10pt}
\end{figure*}

We investigate classification performance of the compared algorithms using LIBSVM~\cite{chang2011libsvm} with 10-fold cross-validation on two binary classification benchmarks\footnote{Rcv1, Kdd and Webspam can be downloaded at\\https://www.csie.ntu.edu.tw/{\texttildelow}cjlin/libsvmtools/datasets/binary.html}. We repeat each experiment 10 times and compute the mean and the standard derivation of results. We adopt the following two data sets (see Table~\ref{tbl:datasets} for details).

\begin{enumerate}
\item \textbf{Rcv1}: The data set is a large collection of newswire stories drawn from online databases. The formatted data set has 20,242 training samples with 47,236 features. The data set has been categorized into two classes: positive instances contain CCAT and ECAT while negative ones contain GCAT and MACT on the website. We randomly select 10,000 positive instances and 10,000 negative ones to compose a balanced data set for classification.
\item \textbf{Kdd}: This is a large educational data set from the KDD Cup 2010 competition. The formatted data set has 8,407,752 training samples with 20,216,830 features. We also randomly select 10,000 positive instances and 10,000 negative ones to form a balanced data set for classification.
\end{enumerate}

\subsubsection{Overall Results}

Figure~\ref{fig:classification} reports the comparison results on Rcv1 and Kdd. Generally speaking, our I$^2$CWS algorithm remains the same classification accuracy as ICWS and [Li,~2015]. On Rcv1, I$^2$CWS outperforms WMH, [Haeupler et.~al.,~2014], and [Gollapudi et.~al.,~2006] with $D$ increasing; while the three of the four CWS algorithms (ICWS, [Li,~2015], I$^2$CWS) perform slightly better than the other one, CCWS. On Kdd, all the algorithms achieve almost the same classification accuracy.

For runtime (only including the time for hashing), the difference between I$^2$CWS and the other three CWS algorithms (ICWS, [Li,~2015], CCWS) is small. As $D$ increases, I$^2$CWS costs slightly more than the other three because the aggregated additional time cost for generating the independent $y_{k_*}$ becomes obvious (Lines 12--13 in Algorithm~\ref{alg:ccws}). By contrast, I$^2$CWS generally runs more efficiently than WMH and [Haeupler et.~al.,~2014] in most cases.
%In this experiment, each algorithm is given a cutoff time of 200 seconds

\begin{figure*}[t]
\centering
\includegraphics[width=\linewidth]{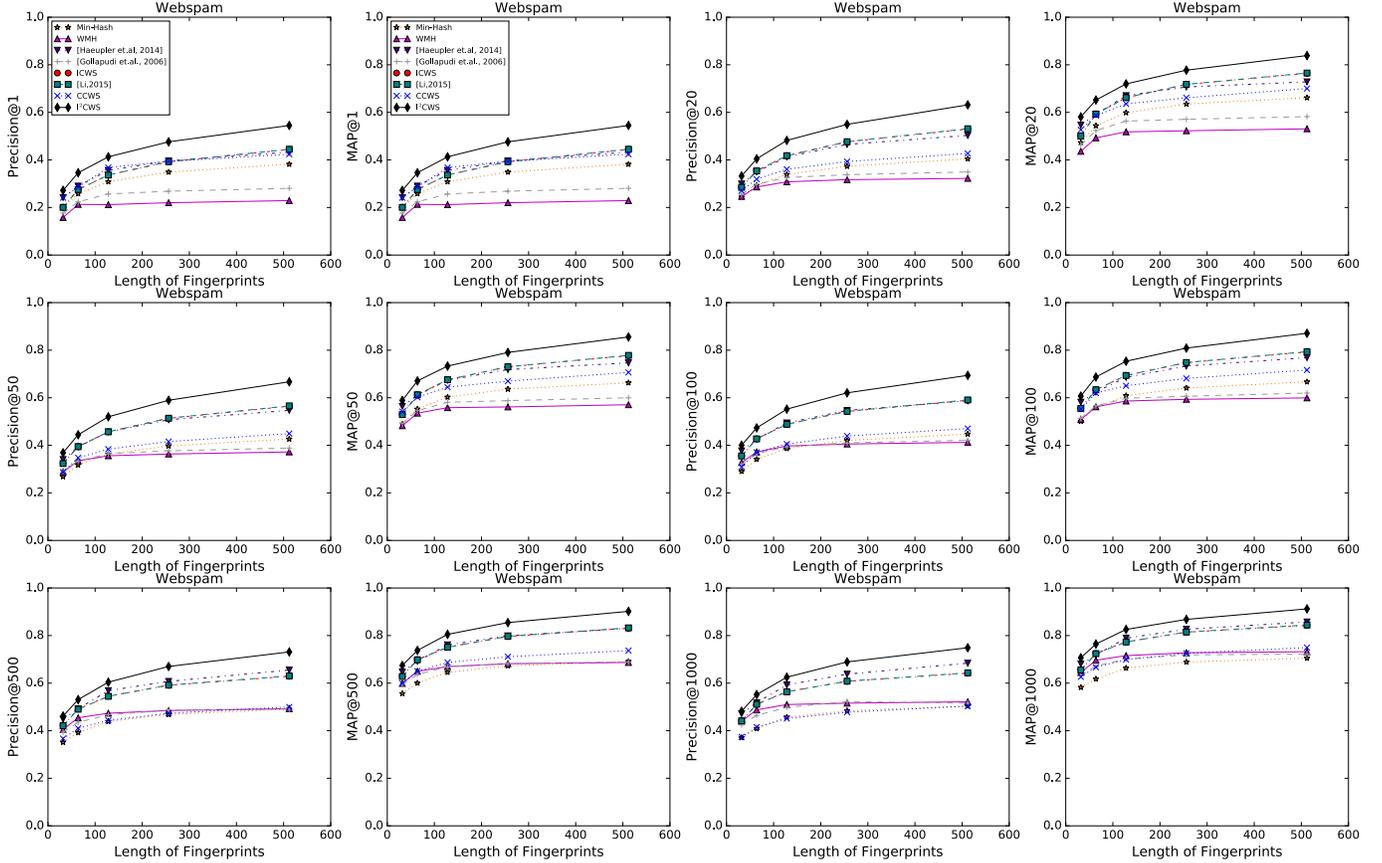}
    \caption{Retrieval results in Precision@$K$ (odd columns) and MAP@$K$ (even columns) of the compared methods on Webspam. The $x$-axis denotes the length of fingerprints, $D$.}
    \label{fig:webspam_performance}
\end{figure*}

\begin{figure*}[t]
\centering
\includegraphics[width=\linewidth]{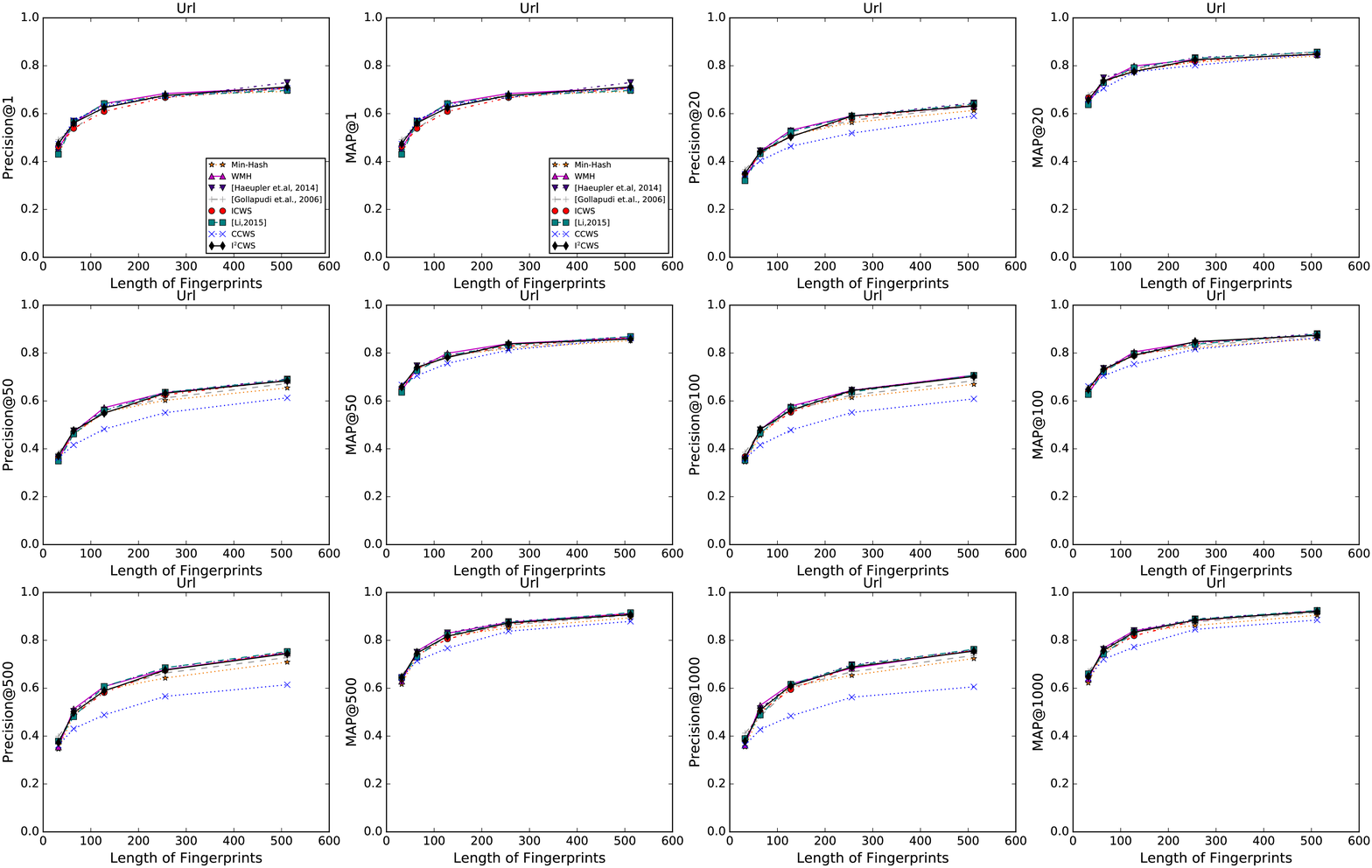}
    \caption{Retrieval results in Precision@$K$ (odd columns) and MAP@$K$ (even columns) of the compared methods on Url. The $x$-axis denotes the length of fingerprints, $D$.}
    \label{fig:url_performance}
\end{figure*}

\subsubsection{Discussion on the Results}

The overall results on Rcv1 and Kdd might not be surprising if we understand the mechanism of the compared algorithms and the statistics of the data sets. I$^2$CWS, ICWS, and CCWS all return the hash code $(k_*, y_{k_*})$ while [Li,~2015] only returns one component, $k_*$, of $(k_*, y_{k_*})$ generated by ICWS. In~\cite{li20150}, it is empirically shown that the contribution of $y_{k_*}$ produced by the CWS scheme is trivial to classification on most data sets; only using $k_*$ as the hash code to produce the fingerprint is sufficient to achieve almost the same classification performance as using $(k_*, y_{k_*})$ -- This phenomenon has been verified again by the above experiments that the three CWS algorithms, I$^2$CWS, ICWS, and [Li,~2015], achieve almost the same performance. The difference among the three CWS algorithms is all about how to sample $y_{k_*}$, which has less empirical contribution, despite a theoretical improvement in I$^2$CWS to relieve the dependence between $k_*$ and $y_{k_*}$.
%Modeled on ICWS, CCWS uniformly discretizes the original weights which avoids discretizing the logarithm of the weights in ICWS; by contrast, our I$^2$CWS algorithm returns the hash code $(k, y_k)$ in a new way of computing $y_k$ that relieves the dependence in ICWS; [Li,~2015] simply discards $y_k$ of $(k, y_k)$ produced by ICWS.

We also notice that CCWS performs worse than the other three CWS algorithms on Rcv1 while it generally competes well on Kdd. CCWS uniformly discretizes the original weights instead of discretizing the logarithm of the weights in ICWS, [Li,~2015], and I$^2$CWS. However, as shown in~\cite{wu2016canonical}, taking logarithm on the weights leads to an increased probability of collision, and such a sublinear (logarithm) transform of the weights is particularly effective for classifying data with large variances of weights (with a similar effect as inverse document frequency), such that the classification performance is likely to be improved. We can see that Rcv1 has a relatively large variance of weights (see Table~\ref{tbl:datasets}) which is more suitable for ICWS, [Li,~2015], and I$^2$CWS.

WMH and [Haeupler et.~al.,~2014] perform slightly worse than the four CWS algorithms on Rcv1 as $D$ increases. There exists a tradeoff related to the scaling factor: If using a larger factor, the performance may be improved while the runtime will be increased because of the dramatically expanded collection of subelements as the universal set. In Kdd (see Table~\ref{tbl:datasets}), each document (weighted set) has around 8,000 non-zero features (positive weights) on average, which is much more than Rcv1, which has about 80 non-zero features. Therefore, we can see that, although the same scaling factor is applied to the two data sets, WMH and [Haeupler et.~al.,~2014] perform more efficiently on Rcv1 than Kdd, because Kdd has a much larger universal set induced by scaling and quantizing.

The reason that all the algorithms maintain the same accuracy level on Kdd may be due to the low variance of the weights, which implies that the weighted set tends to be a binary set.

\subsection{Results on Top-$K$ Retrieval}
\label{subsec:retrieval}

In this experiment, we carry out top-$K$ retrieval, for $K=\{1, 20, 50, 100, 500, 1000\}$. We adopt Precision$@K$ and Mean Average Precision (MAP)$@K$ to measure the performance in terms of accuracy because precision is relatively more important than recall in large-scale retrieval; furthermore, MAP contains information of relative orders of the retrieved samples, which can reflect the retrieval quality more accurately. To this end, we adopt two very large public data sets (see Table~\ref{tbl:datasets} for details).
\begin{enumerate}
  \item \textbf{Webspam}: It is a web text data set provided by a large-scale learning challenge. The data set has 350,000 instances and 16,609,143 features. We randomly select 1,000 samples from the original data set as query examples and the rest as the database.
  \item \textbf{Url}~\cite{ma2009identifying}: The data set contains 2,396,130 URLs and 3,231,961 features. We randomly select 1,000 samples from the original data set as query examples and the rest as the database.
\end{enumerate}

\subsubsection{Overall Results}

Figure~\ref{fig:webspam_performance} and Figure~\ref{fig:url_performance} report the comparison results on Webspam and Url, respectively. On Webspam, our I$^2$CWS algorithm defeat all the other algorithms under all configurations, with clear performance gain. I$^2$CWS outperforms ICWS, [Li,~2015], and [Haeupler et.~al.,~2014] by about 5\% when $D=512$ and performs much better than CCWS. Also, the performance gain of I$^2$CWS over WMH and [Gollapudi et.~al.,~2006] becomes clearer as $D$ increases. By contrast, on Url, all the algorithms except CCWS maintain the same performance level.

\begin{figure}[t]
\centering
\includegraphics[width=\linewidth]{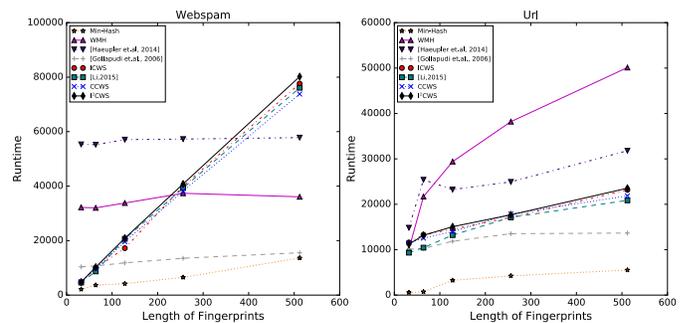}
\caption{Retrieval runtime of the compared methods on Webspam and Url. The $x$-axis denotes the length of fingerprints, $D$.}
\label{fig:webspam_url_time}
%\vspace{-20pt}
\end{figure}

In terms of runtime in Figure~\ref{fig:webspam_url_time}, I$^2$CWS runs slightly slower than the other three CWS algorithms on both data sets.

\subsubsection{Discussion on the Results}

In contrast to that all the CWS algorithms have almost the same classification accuracy, they obtain very different retrieval results on Webspam. The main reason may be due to the effect of the learning process of the classifier -- The similarity computed based on the hash code can be adjusted through the learned coefficients of the classifier to the optimum. Therefore, different algorithms are able to obtain similar classification performance. By contrast, information retrieval totally relies on similarity comparison without being affected by other factors, which may authentically reflect the ability of the hash algorithm.

We notice that our I$^2$CWS algorithm performs much better than the others on Webspam while all the algorithms except CCWS achieve almost the same performance on Url. The difference of the results on the two data sets is mainly caused by the variances of the weights: Webspam has a normal variance of the weights at the same level as those of Rcv1 and Kdd; while Url has an almost vanishing variance such that the data set can be degraded to a binary set to some extent. On Webspam with a normal variance of the weights, I$^2$CWS is able to better approximate the generalized Jaccard similarity by relieving the dependence.

The runtime results again show that all the CWS algorithms share the same time complexity, except for I$^2$CWS costing slightly more for generating $y_{k_*}$. On the other hand, as Webspam is denser than Url, the scaling factor has more impact on Url than Webspam for the expansion of the subelements. Consequently, WMH and [Haeupler et.~al.,~2014] perform more stably on Webspam than Url.

\section{Related Work}
\label{sec:related}

LSH is used to approximate similarity (or distance) measures. So far various LSH methods have been proposed based on $l_p$ distance, angle-based distance, hamming distance and Jaccard similarity, which have been widely applied in statistics \cite{eshghi2008locality}, computer vision \cite{ke2004efficient,zhao2013sim}, multimedia \cite{yu2010combining}, data mining \cite{xiong2015top,yu2017generic}, machine learning \cite{satuluri2012bayesian, neyshabur2015on}, natural language processing \cite{ravichandran2005randomized,ture2011no}, etc. LSH with $p$-stable distribution \cite{datar2004locality} is designed for $l_p$ norm $||\x_i-\x_j||_p$, where $p\in (0,2]$. This scheme employs the property of stable distributions, e.g., Cauchy distribution ($p=1$) and Gaussian distribution ($p=2$), to estimate the corresponding $l_p$ norm. Andoni and Indyk \cite{andoni2006near} extend LSH with $p$-stable distribution in \cite{datar2004locality} into the multi-dimensional space by randomly projecting data points into $\mathbb{R}^t$. Dasgupta et al. \cite{dasgupta2011fast} fast estimate $l_2$ distance between two vectors using randomized Hadamard transforms in a non-linear settting. The Sim-Hash \cite{charikar2002similarity}, as the classical angle-based LSH, is designed to approximate cosine distance between vectors representing data points. In the approach, the vectors are projected into the normal vector of a random hyperplane, and the hash values (0 or 1) are either side of the hyperplane on which the vector lies. Manku et al. \cite{manku2007detecting} practically implement Sim-Hash and propose an algorithmic technique to judge whether a document represented as $D$-bit fingerprints is different from a given document represented as fingerprints with the same bit number in at most $k$ bit-positions of the fingerprints where $k$ is small. Ji. et al. \cite{ji2012super} improve Sim-Hash by partitioning the random projections into different groups and orthogonalizing them in each group. Consequently, their results in each group can be combined together. Kulis et al. \cite{kulis2009kernelized, kulis2012kernelized} extend the angle between vectors in \cite{charikar2002similarity} into the angle between kernel functions, while multiple kernel LSH approaches \cite{wang2010s3mkl,wang2012s3mkl} generate hash functions by adopting multiple kernels, each of which is assigned to the same number of bits. Xia et al. \cite{xia2012boosting} present a boosted version of multi-kernel LSH. Instead of assigning the same number of bits in each kernel, the method in \cite{xia2012boosting} automatically allocates various number of bits via the boosting scheme. The above angle-based LSH approaches are used to retrieve points (or vectors) that are close to a query point (or vectors), while hyperplane hashing aims to retrieve points that are close to a query hyperplane \cite{jain2010hashing,vijayanarasimhan2013Hashing}. For binary vectors, the LSH method for the hamming distance is proposed in \cite{indyk1998approximate}, where one LSH function randomly samples one index from the binary vector. Besides, Gorisse et al. \cite{gorisse2012locality} propose a LSH method for the $\chi^2$ distance between two vectors.

Min-Hash \cite{broder1997resemblance, broder1998min} is proposed to approximate the Jaccard similarity and has been widely applied in the bag-of-words model, for example, duplicate webpage detection \cite{fetterly2003large, henzinger2006finding}, webspam detection \cite{jindal2008opinion,urvoy2008tracking}, text mining \cite{chi2014context, kim2011text}, large-scale machine learning systems \cite{li2012one,li2011hashing}, content matching \cite{pandey2009nearest}, etc. Furthermore, Shrivastava and Li \cite{shrivastava2014defense} give theoretical and experimental evidence that Min-Hash outperforms Sim-Hash in document analysis where data are represented as binary sets. Also, many variations of Min-Hash have been subsequently proposed to improve the efficiency because Min-Hash needs $K$ random permutations. To this end, Min-Max Hash \cite{ji2013min} generates $K$ hash values by employing only $\frac{K}{2}$ random permutations and taking the smallest as well as the largest values of each permutation, but it is still able to obtain an unbiased estimator. Moreover, some methods use only one permutation to speed up Min-Hash. For example, Conditional Random Sampling \cite{li2006conditional, li2007sketch} achieves better estimators than random projections by permutating only once and taking the $k$ smallest nonzeros. By contrast, One Permutation Hashing \cite{li2012one} permutates only once, breaks the permutation into $K$ bins and then concatenates the smallest nonzero value in each bin as a fingerprint. Unfortunately, the method in \cite{li2012one} gives rise to the issue of empty bins, and subsequently, Shrivastava and Li solves the problem and supplies an unbiased estimator in \cite{shrivastava2014densifying}. On the other hand, $b$-bit Min-Hash \cite{li2010b, li2010b2, li2011b} remarkably improves the storage efficiency and provides an unbiased estimator by storing only the lowest $b$ bits of each Min-Hash value (e.g., $b=1$ or $2$) in the case of $K$ permutations. %but also can be integrated with linear learning algorithms, e.g., linear SVM and logistic regression, to solve large-scale and high-dimensional learning tasks.

The aforementioned algorithms regarding the standard Min-Hash are all designed to approximate the Jaccard similarity for binary sets. Subsequently, some weighted Min-Hash algorithms have been proposed to approximate the generalized Jaccard similarity for the weighted sets because the generalized Jaccard similarity is able to sketch sets more accurately than the Jaccard similarity. The naive idea of weighted Min-Hash algorithm is to quantize each weighted element into a number of distinct and equal-sized subelements, and then apply the standard Min-Hash to the collection of subelements. The remaining float part of each weighted element stemming from quantization can be operated by either rounding off or saving with probability \cite{haeupler2014consistent}. Despite the feasibility, the quantization process explicitly increases the size of the universal set which significantly increases the computational workload because each subelement is independently permutated according to the definition of Min-Hash. In order to conquer the issue, the second method in \cite{gollapudi2006exploiting} applies the standard Min-Hash to the binary sets which are transformed by thresholding real-value weights with random samples; by contrast, Chum et al. \cite{chum2008near} compute the Min-Hash values for integer weighted sets by deducing the minimum of a set of random variables. Although the two approaches are efficient -- the second method in \cite{gollapudi2006exploiting} traverses the original set twice and permutates once for every Min-Hash value while the method in \cite{chum2008near} traverses only once for every Min-Hash value, there exist gaps between the expectation of the approximation and the true Jaccard similarity.

In order to further improve the effectiveness and efficiency, researchers have resorted to sampling-based methods. Shrivastava \cite{shrivastava2016simple} proposes a weighted Min-Hash algorithm based on uniform sampling. Unfortunately, it requires to know the upper bound of each element in the universal set in advance, and thus it is not practical in real-world applications. As a milestone work, the first method in \cite{gollapudi2006exploiting} proposes the idea of ``active indices". The ``active indices" are independently sampled on a weighted element as a sequence of subelements whose hash values are monotonically decreasing. Consequently, a large number of inactive subelements between two adjacent ``active indices" are skipped, and the computational complexity is proportional to the logarithm of the weight. However, this method is still inefficient for real-value weighted sets because real-value weights must be transformed into integer weights by multiplying a large constant. Subsequently, the CWS scheme \cite{manasse2010consistent} is proposed to solve the efficiency problem for real-value weighted sets by considering only two special ``active indices", one of which is the largest ``active indices" smaller than the weight and the other of which is the smallest ``active indices" greater than the weight. The algorithm in \cite{manasse2010consistent} still needs to traverse some ``active indices" in order to find the two special ones, and thus it runs in expected constant time for weighted elements; while ICWS \cite{ioffe2010improved} runs in worst-case constant time by directly sampling the two special ones. Recently, Wu et al. \cite{wu2017consistent} uncovers the working mechanism of ICWS -- ICWS essentially needs to sample five independent uniform random variables for each element -- and further proposes Practical CWS (PCWS) which is simpler and more efficient in terms of time and space complexities by sampling four independent uniform random variables for each element. Li \cite{li20150} approximates ICWS by simply discarding one component of the Min-Hash values returned by ICWS, and empirically shows that the discarded component hardly affects performance. However, Wu et al. \cite{wu2016canonical} claims that the three CWS algorithms (ICWS, PCWS and the algorithm in \cite{li20150}) all conduct uniform discretization on the logarithm of the weight, and thus there is a risk of violating the uniformity of the CWS scheme to some extent. In order to avoid the risk, Wu et al. \cite{wu2016canonical} proposes Canonical CWS (CCWS) by uniformly discretizing the original weights, but it decreases the probability of collision and thus generally performs worse than the three other CWS algorithms.

 %Essentially, ICWS needs to sample five independent uniform random variables for each element, while Practical CWS (PCWS) \cite{wu2017consistent} uncovers the working mechanism of ICWS, and thus makes it simpler and more efficient in terms of time and space complexities by sampling four independent uniform random variables for each element. Recently, \cite{li20150} approximates ICWS by simply discarding one component of the Min-Hash values returned by ICWS, and empirically shows that the discarded component hardly affects performance. However, the three CWS algorithms (ICWS, PCWS and \cite{li20150}) all conduct uniform discretization on the logarithm of the weight. As a result, there is a risk of violating the uniformity of the CWS scheme to some extent; by contrast, Canonical CWS (CCWS) \cite{wu2016canonical} uniformly discretizes the original weights in order to avoid the issue, but it decreases the probability of collision and thus generally performs worse than the three CWS algorithms.

\section{Conclusion}
\label{sec:con}

In this paper, we propose the Improved ICWS (I$^2$CWS) algorithm to relieve the underlying dependence between the two components of the hash code produced by, ICWS~\cite{ioffe2010improved}, the widely accepted state-of-the-art for real-value weighted Min-Hash. The proposed I$^2$CWS algorithm not only complies with the CWS scheme but also shares the same computational complexity as ICWS. We conduct extensive empirical tests of the proposed I$^2$CWS algorithm and the state-of-the-arts for estimating the generalized Jaccard similarity on two groups of synthetic data sets, as well as for classification and information retrieval on four real-world text data sets. The experimental results demonstrate that I$^2$CWS is able to estimate the generalized Jaccard similarity more accurately than ICWS, and furthermore compete with or outperform the compared methods while keeping the similar efficiency as ICWS.

According to our empirical tests, we have the following intersting findings: 1) $y_{k_*}$ produced by the CWS scheme indeed has less contribution to classification performance, which has been observed in~\cite{li20150}. 2) The classification results of all the CWS algorithms are similar which may be due to the effect of learning process;  information retrieval should be a more proper task for evaluating weighted Min-Hash algorithms. 3) I$^2$CWS is indeed more effective for approximating the generalized Jaccard similarity by relieving the underlying dependence.

\ifCLASSOPTIONcaptionsoff
  \newpage
\fi

% trigger a \newpage just before the given reference
% number - used to balance the columns on the last page
% adjust value as needed - may need to be readjusted if
% the document is modified later
%\IEEEtriggeratref{8}
% The "triggered" command can be changed if desired:
%\IEEEtriggercmd{\enlargethispage{-5in}}

% references section

% can use a bibliography generated by BibTeX as a .bbl file
% BibTeX documentation can be easily obtained at:
% http://mirror.ctan.org/biblio/bibtex/contrib/doc/
% The IEEEtran BibTeX style support page is at:
% http://www.michaelshell.org/tex/ieeetran/bibtex/
%\bibliographystyle{IEEEtran}
% argument is your BibTeX string definitions and bibliography database(s)
%\bibliography{IEEEabrv,../bib/paper}
%
% <OR> manually copy in the resultant .bbl file
% set second argument of \begin to the number of references
% (used to reserve space for the reference number labels box)

% biography section
%
% If you have an EPS/PDF photo (graphicx package needed) extra braces are
% needed around the contents of the optional argument to biography to prevent
% the LaTeX parser from getting confused when it sees the complicated
% \includegraphics command within an optional argument. (You could create
% your own custom macro containing the \includegraphics command to make things
% simpler here.)
%\begin{IEEEbiography}[{\includegraphics[width=1in,height=1.25in,clip,keepaspectratio]{mshell}}]{Michael Shell}
% or if you just want to reserve a space for a photo:
\bibliographystyle{IEEEtran}
\bibliography{main}

% Generated by IEEEtran.bst, version: 1.13 (2008/09/30)
\begin{thebibliography}{10}
\providecommand{\url}[1]{#1}
\csname url@samestyle\endcsname
\providecommand{\newblock}{\relax}
\providecommand{\bibinfo}[2]{#2}
\providecommand{\BIBentrySTDinterwordspacing}{\spaceskip=0pt\relax}
\providecommand{\BIBentryALTinterwordstretchfactor}{4}
\providecommand{\BIBentryALTinterwordspacing}{\spaceskip=\fontdimen2\font plus
\BIBentryALTinterwordstretchfactor\fontdimen3\font minus
  \fontdimen4\font\relax}
\providecommand{\BIBforeignlanguage}[2]{{%
\expandafter\ifx\csname l@#1\endcsname\relax
\typeout{** WARNING: IEEEtran.bst: No hyphenation pattern has been}%
\typeout{** loaded for the language `#1'. Using the pattern for}%
\typeout{** the default language instead.}%
\else
\language=\csname l@#1\endcsname
\fi
#2}}
\providecommand{\BIBdecl}{\relax}
\BIBdecl

\bibitem{google}
D.~Sullivan, \emph{Google now handles at least 2 trillion searches per year},
  May 2016,
  http://searchengineland.com/google-now-handles-2-999-trillion-searches-per-year-250247.

\bibitem{facebook}
L.~Goode, \emph{Messenger and WhatsApp process 60 billion messages a day, three
  times more than SMS}, April 2016,
  http://www.theverge.com/2016/4/12/11415198/facebook-messenger-whatsapp-number-messages-vs-sms-f8-2016.

\bibitem{dumbill2013revolution}
E.~Dumbill, ``{A} {R}evolution {T}hat {W}ill {T}ransform {H}ow {W}e {L}ive,
  {W}ork, and {T}hink: {A}n {I}nterview with the {A}uthors of {B}ig {D}ata,''
  \emph{Big Data}, vol.~1, no.~2, pp. 73--77, 2013.

\bibitem{rajaraman2012mining}
A.~Rajaraman, J.~D. Ullman, J.~D. Ullman, and J.~D. Ullman, \emph{{M}ining of
  {M}assive {D}atasets}.\hskip 1em plus 0.5em minus 0.4em\relax Cambridge
  University Press Cambridge, 2012, vol.~1.

\bibitem{indyk1998approximate}
P.~Indyk and R.~Motwani, ``{A}pproximate {N}earest {N}eighbors: {T}owards
  {R}emoving the {C}urse of {D}imensionality,'' in \emph{STOC}, 1998, pp.
  604--613.

\bibitem{gionis1999similarity}
A.~Gionis, P.~Indyk, R.~Motwani \emph{et~al.}, ``{S}imilarity {S}earch in
  {H}igh {D}imensions via {H}ashing,'' in \emph{VLDB}, vol.~99, no.~6, 1999,
  pp. 518--529.

\bibitem{broder1998min}
A.~Z. Broder, M.~Charikar, A.~M. Frieze, and M.~Mitzenmacher, ``{M}in-wise
  {I}ndependent {P}ermutations,'' in \emph{STOC}, 1998, pp. 327--336.

\bibitem{charikar2002similarity}
M.~S. Charikar, ``{S}imilarity {E}stimation {T}echniques from {R}ounding
  {A}lgorithms,'' in \emph{STOC}, 2002, pp. 380--388.

\bibitem{manku2007detecting}
G.~S. Manku, A.~Jain, and A.~Das~Sarma, ``{D}etecting {N}ear-{D}uplicates for
  {W}eb {C}rawling,'' in \emph{WWW}, 2007, pp. 141--150.

\bibitem{datar2004locality}
M.~Datar, N.~Immorlica, P.~Indyk, and V.~S. Mirrokni, ``{L}ocality-{S}ensitive
  {H}ashing {S}cheme {B}ased on p-{S}table {S}istributions,'' in \emph{SOCG},
  2004, pp. 253--262.

\bibitem{shrivastava2014defense}
A.~Shrivastava and P.~Li, ``{I}n {D}efense of {M}inhash {O}ver {S}im{H}ash,''
  in \emph{AISTATS}, 2014, pp. 886--894.

\bibitem{li2010b}
P.~Li and C.~K{\"o}nig, ``$b$-{B}it {M}inwise {H}ashing,'' in \emph{WWW}, 2010,
  pp. 671--680.

\bibitem{li2012one}
P.~Li, A.~Owen, and C.-H. Zhang, ``{O}ne {P}ermutation {H}ashing,'' in
  \emph{NIPS}, 2012, pp. 3113--3121.

\bibitem{mitzenmacher2014efficient}
M.~Mitzenmacher, R.~Pagh, and N.~Pham, ``{E}fficient {E}stimation for {H}igh
  {S}imilarities {U}sing {O}dd {S}ketches,'' in \emph{WWW}, 2014, pp. 109--118.

\bibitem{shrivastava2014densifying}
A.~Shrivastava and P.~Li, ``{D}ensifying {O}ne {P}ermutation {H}ashing via
  {R}otation for {F}ast {N}ear {N}eighbor {S}earch,'' in \emph{ICML}, 2014, pp.
  557--565.

\bibitem{haveliwala2000scalable}
T.~H. Haveliwala, A.~Gionis, and P.~Indyk, ``{S}calable {T}echniques for
  {C}lustering the {W}eb,'' in \emph{WebDB}, 2000, pp. 129--134.

\bibitem{haeupler2014consistent}
B.~Haeupler, M.~Manasse, and K.~Talwar, ``{C}onsistent {W}eighted {S}ampling
  {M}ade {F}ast, {S}mall, and {E}asy,'' \emph{arXiv preprint arXiv:1410.4266},
  2014.

\bibitem{gollapudi2006exploiting}
S.~Gollapudi and R.~Panigrahy, ``{E}xploiting {A}symmetry in {H}ierarchical
  {T}opic {E}xtraction,'' in \emph{CIKM}, 2006, pp. 475--482.

\bibitem{manasse2010consistent}
M.~Manasse, F.~McSherry, and K.~Talwar, ``{C}onsistent {W}eighted {S}ampling,''
  \emph{Unpublished technical report}, 2010.

\bibitem{ioffe2010improved}
S.~Ioffe, ``{I}mproved {C}onsistent {S}ampling, {W}eighted {M}inhash and {L}$1$
  {S}ketching,'' in \emph{ICDM}, 2010, pp. 246--255.

\bibitem{wu2017consistent}
W.~Wu, B.~Li, L.~Chen, and C.~Zhang, ``{C}onsistent {W}eighted {S}ampling
  {M}ade {M}ore {P}ractical,'' in \emph{WWW}, 2017, pp. 1035--1043.

\bibitem{li20150}
P.~Li, ``$0$-{B}it {C}onsistent {W}eighted {S}ampling,'' in \emph{KDD}, 2015,
  pp. 665--674.

\bibitem{wu2016canonical}
W.~Wu, B.~Li, L.~Chen, and C.~Zhang, ``{C}anonical {C}onsistent {W}eighted
  {S}ampling for {R}eal-{V}alue {W}eighted {M}in-{H}ash,'' in \emph{ICDM},
  2016, pp. 1287--1292.

\bibitem{chang2011libsvm}
C.-C. Chang and C.-J. Lin, ``{LIBSVM}: {A} {L}ibrary for {S}upport {V}ector
  {M}achines,'' \emph{ACM Transactions on Intelligent Systems and Technology},
  vol.~2, no.~3, p.~27, 2011.

\bibitem{ma2009identifying}
J.~Ma, L.~K. Saul, S.~Savage, and G.~M. Voelker, ``{I}dentifying {S}uspicious
  {URL}s: {A}n {A}pplication of {L}arge-{S}cale {O}nline {L}earning,'' in
  \emph{ICML}, 2009, pp. 681--688.

\bibitem{eshghi2008locality}
K.~Eshghi and S.~Rajaram, ``{L}ocality {S}ensitive {H}ash {F}unctions {B}ased
  on {C}oncomitant {R}ank {O}rder {S}tatistics,'' in \emph{KDD}.\hskip 1em plus
  0.5em minus 0.4em\relax ACM, 2008, pp. 221--229.

\bibitem{ke2004efficient}
Y.~Ke, R.~Sukthankar, L.~Huston, Y.~Ke, and R.~Sukthankar, ``{E}fficient
  {N}ear-{D}uplicate {D}etection and {S}ub-{I}mage {R}etrieval,'' in
  \emph{ACMMM}, vol.~4, no.~1, 2004, p.~5.

\bibitem{zhao2013sim}
W.-L. Zhao, H.~J{\'e}gou, and G.~Gravier, ``{S}im-{M}in-{H}ash: {A}n
  {E}fficient {M}atching {T}echnique for {L}inking {L}arge {I}mage
  {C}ollections,'' in \emph{ACMMM}, 2013, pp. 577--580.

\bibitem{yu2010combining}
Y.~Yu, M.~Crucianu, V.~Oria, and E.~Damiani, ``{C}ombining {M}ulti-{P}robe
  {H}istogram and {O}rder-{S}tatistics {B}ased {LSH} for {S}calable {A}udio
  {C}ontent {R}etrieval,'' in \emph{ACMMM}, 2010, pp. 381--390.

\bibitem{xiong2015top}
Y.~Xiong, Y.~Zhu, and S.~Y. Philip, ``{T}op-{K} {S}imilarity {J}oin in
  {H}eterogeneous {I}nformation {N}etworks,'' \emph{IEEE Transactions on
  Knowledge and Data Engineering}, vol.~27, no.~6, pp. 1710--1723, 2015.

\bibitem{yu2017generic}
C.~Yu, S.~Nutanong, H.~Li, C.~Wang, and X.~Yuan, ``{A} {G}eneric {M}ethod for
  {A}ccelerating {LSH}-{B}ased {S}imilarity {J}oin {P}rocessing,'' \emph{IEEE
  Transactions on Knowledge and Data Engineering}, vol.~29, no.~4, pp.
  712--726, 2017.

\bibitem{satuluri2012bayesian}
V.~Satuluri and S.~Parthasarathy, ``{B}ayesian {L}ocality {S}ensitive {H}ashing
  for {F}ast {S}imilarity {S}earch,'' \emph{VLDB}, vol.~5, no.~5, pp. 430--441,
  2012.

\bibitem{neyshabur2015on}
B.~Neyshabur and N.~Srebro, ``{O}n {S}ymmetric and {A}symmetric {LSH}s for
  {I}nner {P}roduct {S}earch,'' in \emph{ICML}, 2015, pp. 1926--1934.

\bibitem{ravichandran2005randomized}
D.~Ravichandran, P.~Pantel, and E.~Hovy, ``{R}andomized {A}lgorithms and {NLP}:
  {U}sing {L}ocality {S}ensitive {H}ash {F}unction for {H}igh {S}peed {N}oun
  {C}lustering,'' in \emph{ACL}, 2005, pp. 622--629.

\bibitem{ture2011no}
F.~Ture, T.~Elsayed, and J.~Lin, ``{N}o {F}ree {L}unch: {B}rute {F}orce vs.
  {L}ocality-{S}ensitive {H}ashing for {C}ross-{L}ingual {P}airwise
  {S}imilarity,'' in \emph{SIGIR}, 2011, pp. 943--952.

\bibitem{andoni2006near}
A.~Andoni and P.~Indyk, ``{N}ear-{O}ptimal {H}ashing {A}lgorithms for
  {A}pproximate {N}earest {N}eighbor in {H}igh {D}imensions,'' in \emph{FOCS},
  2006, pp. 459--468.

\bibitem{dasgupta2011fast}
A.~Dasgupta, R.~Kumar, and T.~Sarl{\'o}s, ``{F}ast {L}ocality-{S}ensitive
  {H}ashing,'' in \emph{KDD}, 2011, pp. 1073--1081.

\bibitem{ji2012super}
J.~Ji, J.~Li, S.~Yan, B.~Zhang, and Q.~Tian, ``{S}uper-{B}it
  {L}ocality-{S}ensitive {H}ashing,'' in \emph{NIPS}, 2012, pp. 108--116.

\bibitem{kulis2009kernelized}
B.~Kulis and K.~Grauman, ``{K}ernelized {L}ocality-{S}ensitive {H}ashing for
  {S}calable {I}mage {S}earch,'' in \emph{ICCV}, 2009, pp. 2130--2137.

\bibitem{kulis2012kernelized}
------, ``{K}ernelized {L}ocality-{S}ensitive {H}ashing,'' \emph{IEEE
  Transactions on Pattern Analysis and Machine Intelligence}, vol.~34, no.~6,
  pp. 1092--1104, 2012.

\bibitem{wang2010s3mkl}
S.~Wang, S.~Jiang, Q.~Huang, and Q.~Tian, ``{S$^3$MKL}: {S}calable
  {S}emi-{S}upervised {M}ultiple {K}ernel {L}earning for {I}mage {D}ata
  {M}ining,'' in \emph{ACMMM}, 2010, pp. 163--172.

\bibitem{wang2012s3mkl}
S.~Wang, Q.~Huang, S.~Jiang, and Q.~Tian, ``{S}$^{3}${MKL}: {S}calable
  {S}emi-{S}upervised {M}ultiple {K}ernel {L}earning for {R}eal-{W}orld {I}mage
  {A}pplications,'' \emph{IEEE Transactions on Multimedia}, vol.~14, no.~4, pp.
  1259--1274, Aug 2012.

\bibitem{xia2012boosting}
H.~Xia, P.~Wu, S.~C. Hoi, and R.~Jin, ``{B}oosting {M}ulti-{K}ernel
  {L}ocality-{S}ensitive {H}ashing for {S}calable {I}mage {R}etrieval,'' in
  \emph{SIGIR}, 2012, pp. 55--64.

\bibitem{jain2010hashing}
P.~Jain, S.~Vijayanarasimhan, and K.~Grauman, ``{H}ashing {H}yperplane
  {Q}ueries to {N}ear {P}oints with {A}pplications to {L}arge-{S}cale {A}ctive
  {L}earning,'' in \emph{NIPS}, 2010, pp. 928--936.

\bibitem{vijayanarasimhan2013Hashing}
S.~Vijayanarasimhan, P.~Jain, and K.~Grauman, ``{H}ashing {H}yperplane
  {Q}ueries to {N}ear {P}oints with {A}pplications to {L}arge-{S}cale {A}ctive
  {L}earning,'' \emph{IEEE Transactions on Pattern Analysis and Machine
  Intelligence}, vol.~36, no.~2, pp. 276--288, Feb 2014.

\bibitem{gorisse2012locality}
D.~Gorisse, M.~Cord, and F.~Precioso, ``{L}ocality-{S}ensitive {H}ashing for
  {C}hi2 {D}istance,'' \emph{IEEE Transactions on Pattern Analysis and Machine
  Intelligence}, vol.~34, no.~2, pp. 402--409, 2012.

\bibitem{broder1997resemblance}
A.~Z. Broder, ``{O}n the {R}esemblance and {C}ontainment of {D}ocuments,'' in
  \emph{Compression and Complexity of Sequences 1997. Proceedings}, 1997, pp.
  21--29.

\bibitem{fetterly2003large}
D.~Fetterly, M.~Manasse, M.~Najork, and J.~Wiener, ``{A} {L}arge-{S}cale
  {S}tudy of the {E}volution of {W}eb {P}ages,'' in \emph{WWW}, 2003, pp.
  669--678.

\bibitem{henzinger2006finding}
M.~Henzinger, ``{F}inding {N}ear-{D}uplicate {W}eb {P}ages: {A} {L}arge-{S}cale
  {E}valuation of {A}lgorithms,'' in \emph{SIGIR}, 2006, pp. 284--291.

\bibitem{jindal2008opinion}
N.~Jindal and B.~Liu, ``{O}pinion {S}pam and {A}nalysis,'' in \emph{WSDM},
  2008, pp. 219--230.

\bibitem{urvoy2008tracking}
T.~Urvoy, E.~Chauveau, P.~Filoche, and T.~Lavergne, ``{T}racking {W}eb {S}pam
  with {HTML} {S}tyle {S}imilarities,'' \emph{ACM Transactions on the Web
  (TWEB)}, vol.~2, no.~1, p.~3, 2008.

\bibitem{chi2014context}
L.~Chi, B.~Li, and X.~Zhu, ``{C}ontext-{P}reserving {H}ashing for {F}ast {T}ext
  {C}lassification.'' in \emph{SDM}, 2014, pp. 100--108.

\bibitem{kim2011text}
C.~Kim and K.~Shim, ``{T}ext: {A}utomatic {T}emplate {E}xtraction from
  {H}eterogeneous {W}eb {P}ages,'' \emph{IEEE Transactions on Knowledge and
  Data Engineering}, vol.~23, no.~4, pp. 612--626, 2011.

\bibitem{li2011hashing}
P.~Li, A.~Shrivastava, J.~L. Moore, and A.~C. K{\"o}nig, ``{H}ashing
  {A}lgorithms for {L}arge-{S}cale {L}earning,'' in \emph{NIPS}, 2011, pp.
  2672--2680.

\bibitem{pandey2009nearest}
S.~Pandey, A.~Broder, F.~Chierichetti, V.~Josifovski, R.~Kumar, and
  S.~Vassilvitskii, ``{N}earest-{N}eighbor {C}aching for {C}ontent-{M}atch
  {A}pplications,'' in \emph{WWW}, 2009, pp. 441--450.

\bibitem{ji2013min}
J.~Ji, J.~Li, S.~Yan, Q.~Tian, and B.~Zhang, ``{M}in-{M}ax {H}ash for {J}accard
  {S}imilarity,'' in \emph{ICDM}, 2013, pp. 301--309.

\bibitem{li2006conditional}
P.~Li, K.~W. Church, and T.~Hastie, ``{C}onditional {R}andom {S}ampling: {A}
  {S}ketch-based {S}ampling {T}echnique for {S}parse {D}ata,'' in \emph{NIPS},
  2006, pp. 873--880.

\bibitem{li2007sketch}
P.~Li and K.~W. Church, ``{A} {S}ketch {A}lgorithm for {E}stimating {T}wo-{W}ay
  and {M}ulti-{W}ay {A}ssociations,'' \emph{Computational Linguistics},
  vol.~33, no.~3, pp. 305--354, 2007.

\bibitem{li2010b2}
P.~Li, A.~Konig, and W.~Gui, ``b-{B}it {M}inwise {H}ashing for {E}stimating
  {T}hree-{W}ay {S}imilarities,'' in \emph{NIPS}, 2010, pp. 1387--1395.

\bibitem{li2011b}
P.~Li, A.~Shrivastava, J.~Moore, and A.~C. K{\"o}nig, ``b-{B}it {M}inwise
  {H}ashing for {L}arge-{S}cale {L}earning,'' in \emph{Big Learning 2011: NIPS
  2011 Workshop on Algorithms, Systems, and Tools for Learning at Scale},
  December 2011.

\bibitem{chum2008near}
O.~Chum, J.~Philbin, A.~Zisserman \emph{et~al.}, ``{N}ear {D}uplicate {I}mage
  {D}etection: {M}in-{H}ash and {T}f-{I}df {W}eighting.'' in \emph{BMVC}, vol.
  810, 2008, pp. 812--815.

\bibitem{shrivastava2016simple}
A.~Shrivastava, ``{S}imple and {E}fficient {W}eighted {M}inwise {H}ashing,'' in
  \emph{NIPS}, 2016, pp. 1498--1506.

\end{thebibliography}

\end{document}